\documentclass[epj]{svjour}
\usepackage{xcolor}
\usepackage{graphicx}
\usepackage{caption}
\usepackage{subcaption}  
\usepackage{amsmath}
\usepackage{url}
\usepackage[utf8]{inputenc}  
\usepackage[T1]{fontenc} 
\usepackage{booktabs}
\makeatletter
\let\@afterindentfalse\@afterindenttrue
\makeatother
\title{Excitation function for $^{\boldsymbol{nat}}$Mo(p,x) reactions with covariance analysis
} 

\author{Sumit Bamal\inst{1} \and
S. Lawitlang\inst{2} \and
B. Lalremruata\inst{2} \and
A. Mazumdar\inst{3} \and
S. Pal\inst{4} \and
M. S. Pose\inst{5} \and
V. Nanal\inst{5} \thanks{\email{nanal@tifr.res.in}} \and
Rebecca Pachuau\inst{1} \thanks{\email{pcrl.bec@gmail.com}}}

\institute{Department of Physics, Banaras Hindu University, Varanasi-221005, INDIA \and
Department of Physics, Mizoram University, Aizawl - 796004, INDIA \and
INO, Tata Institute of Fundamental Research, Mumbai - 400005, INDIA \thanks{Presently at
Department of Physics and Astronomy, University of North Carolina, Chapel Hill, NC 27599,
USA} \and
PLF, Tata Institute of Fundamental Research, Mumbai - 400005, INDIA \and
DNAP, Tata Institute of Fundamental Research, Mumbai - 400005, INDIA}

\begin{document}

\date{Received: date / Revised version: date}

\abstract{The excitation functions of proton induced reactions on $^{nat}$Mo 
targets are measured using the activation technique followed by off-line 
$\gamma$-ray spectroscopy with improved precision. The experiment was
performed at the BARC-TIFR Pelletron Linac Facility, Mumbai. Thin samples of  $^{nat}$Mo  were irradiated with a
proton beam of energies ranging from 13 to 22 MeV for measurements of various (p,x) cross sections, with an emphasis on the production of medical isotopes.
The present data for $^{93m}$Tc and $^{94g}$Tc address and resolve the discrepancies in the existing data. 
The $^{93\text{g}}$Tc radioisotope production is extracted with appropriate corrections for $^{93\text{m}}$Tc contribution.
While the measured cross sections for the production of $^{95g}$Tc, $^{96g+m}$Tc, $^{99m}$Tc and $^{99}$Mo are consistent with the reported data, 
significant differences are observed for $^{89g+m}$Zr. 
A comprehensive uncertainty 
analysis, including the correlation coefficients of the measured
reaction cross sections is also presented. \\
\PACS{ 25.40.Lw, 25.40.-h,  24.50.+g  }
}

\maketitle

\section{Introduction}
\label{intro}
Proton induced reactions are of great importance in
various fields including nuclear medicine~\cite{nuclmed}, theranostics applications~\cite{wang2022production}, astrophysical p-processes ~\cite{mazumdar2019studies} and as variable energy neutron sources for research purposes.
The importance of radioisotopes 
for theranostics in modern healthcare cannot be overemphasized. 
Presently, there are around 200 artificially 
 produced medical isotopes, which are 
used in various medical procedures ~\cite{wang2022production}. Innovative ways for 
medical isotope manufacturing are also being studied, including laser isotope separation and 
photo plasma creation~\cite{lobok2022laser}. 
proton induced reactions on Molybdenum are of interest for the 
production of radioisotopes for a wide variety of medical applications such as detection of 
 cancer, heart disease, kidney disease, bone 
disorders, salivary gland disorders, thyroid disorders, PET (Positron Emission Tomography) scans
and SPECT (Single Photon Emission Computed Tomography) scans. 

A range of radioisotopes  $^{99m}$Tc, 
$^{99}$Mo, $^{96g}$Tc, $^{95g}$Tc, $^{94g}$Tc, and $^{89g}$Zr can be produced in $^{nat}$Mo(p,x) reactions. Among these,
the $^{99m}$Tc holds significant importance as it alone accounts for 
approximately 80\% of all nuclear medicinal procedures. The $^{99m}$Tc finds
extensive application in SPECT imaging, particularly in diagnosing strokes. 
The ideal properties of $^{99m}$Tc include its appropriate half-life, penetrating strength,
and minimal biological damage from administered doses, making it a valuable
radio-pharmaceutical for various diagnostic purposes. Further, its nuclear properties also
make it ideal for medical imaging due to its relatively high photon energy ($\approx$140 keV), which allows a precise determination of molecular structure using 
scintillation devices like gamma cameras~\cite{hasan2020molybdenum}.

Given the critical role of medical radioisotope $^{99m}$Tc, there is a wide interest for potential alternatives. In this 
context, Takehito Hayakawa et al.~\cite{hayakawa201895gtc} have
proposed that the radioisotopes $^{96g}$Tc and $^{95g}$Tc could serve 
as viable substitutes for $^{99m}$Tc as $\gamma$-ray emitters.
The long-lived positron emitter $^{89g}$Zr (T$_{1/2}$= 78.41 hrs) is significant in immuno-PET 
scans due to its favorable physical characteristics~\cite{wang2022production}. The $^{94g}$Tc
can be used in PET imaging technique as a carrier-free radionuclide. 
Accurate 
cross section data are essential for these radioisotopes for proposed applications in medical diagnostics.~\cite{khandaker2007measurement}. Production of $^{99}$Mo has traditionally relied on reactor-based routes, either via neutron 
capture on $^{98}$Mo ($^{98}$Mo(n,$\gamma$)$^{99}$Mo) or as a fission product in highly enriched
uranium (HEU) targets. However, increasing restrictions on the use of HEU and the growing global
demand for $^{99\text{m}}$Tc have motivated the development of accelerator based production 
 strategies~\cite{jasim2024proton}, employing protons, deuterons, or electron beams on molybdenum 
targets. Hence, accelerator based production of $^{99}$Mo offers a sustainable and reactor independent 
alternative that is directly relevant to nuclear medicine supply security. The radionuclide 
$^{99}$Mo decays via $\beta^{-}$ emission to produce the clinically important $^{99\text{m}}$Tc,
making it a convenient parent for the on-demand $^{99\text{m}}$Tc generators. 
It should be noted that  cross-disciplinary research not only advances medical science, but also yields important inputs for nuclear reaction mechanism.

Further, Mo(p,x) data may be important for nuclear technology applications. With a melting point of 2623 °C and thermal conductivity of ~112 W/m·K at 1000 K,
molybdenum is compatible with both water and liquid-metal coolants. As a result, it
is suitable for advanced fuel element structures. Relative to zirconium, it reduces
the thermal time constant of fuel rods by nearly a factor of two, which improves 
reactivity feedback and safety margins 
~\cite{shmelev2016kerntechnik,shmelev2017molybdenum}. In dispersion fuels, U–Mo 
alloy granules in a molybdenum matrix provide high fissile density and efficient 
heat transfer, maintaining stability at high burn-up ~\cite{shmelev2017molybdenum}.

Precise determination of nuclear reaction cross sections and incident proton energies is 
essential for improving the reliability of experimental results and for benchmarking 
theoretical models. Low uncertainty in these quantities enables accurate comparison 
between experimental data and model calculations, which is particularly important for 
applications such as medical isotope production and nuclear data evaluations. In addition 
to minimizing statistical and systematic uncertainties, it is important to include 
covariance and correlation information when reporting cross section data. This becomes 
especially relevant when the data are used for interpolation or extrapolation across 
energy ranges where measurements are not directly available. Covariance matrices provide a 
quantitative representation of how uncertainties at different energy points are related.
Ignoring these correlations can lead to incorrect uncertainty propagation and potentially 
misleading results, especially when integrated cross sections or model fits are involved. 
Therefore, including detailed uncertainty analysis along with covariance and correlation 
data significantly enhances the utility and reliability of nuclear reaction data for both 
experimental and theoretical purposes. 
However, the existing datasets in the EXFOR database are limited and show some discrepancies.
Moreover, most of the data is from thick target irradiation and consequently have a large spread in
incident beam energy. Hence, there is a need for additional measurements with high precision, which can be achieved with thin targets. 
 With this motivation, in the present work, 
 the excitation functions for various radioisotopes produced in $^{nat}$Mo(p,x) reactions are measured over an incident beam energy range of 12 to 22 MeV with improved precision. In addition, a detailed uncertainty propagation 
and covariance analysis is also carried out. 
The correlation coefficients between incident proton energy and cross section of nuclear
reactions are not only useful for optimization of production of isotope of interest, but can also provide  insights into reaction mechanisms. 

This paper is organised as follows. The experimental details are described in section 2, analysis
procedure is given in section 3, the results on production of various isotopes are given in section
4 and the conclusions are given in the last section. 

\section{Experimental details}
\label{sec:2}
The experiment was performed at 14UD BARC-TIFR Pelletron Linac Facility in Mumbai, where  high
purity ($\approx$99.9\%) natural molybdenum foils were irradiated with  the proton beams of 
energy  $\sim$ 12 to 22 MeV.  
The thickness of Mo targets was in the range 
 7.79 - 10.34 mg/cm$^2$. For efficient use of beam time, the irradiation was combined with another experiment and a stack of Zr-Cu-Mo foils was used in each run.
 Average thickness of the individual target foils
was determined by measuring its mass and area, assuming uniformity.   The average beam energy in 
the lab frame, at the center of each Mo target foil, and the spread in the energy due to target
thickness were calculated using the SRIM code~\cite{ziegler2003stopping}. The uncertainty in the
incident beam energy, including effects of cascaded targets, is $\sim0.060-0.085$ MeV. Details of
irradiation are given in Table~1. 

\begin{table}[htbp!]
\centering
\caption {Details of irradiation : Proton energy (E$_p$) in center of mass frame at the center of the Mo target, target thickness,  irradiation time (T$_{irr}$) and  average beam current (I).}
\label{tab:1}
\begin{tabular}{lllll}
\hline\noalign{\smallskip}
E$_{p}$  & & Mo Thickness &  T$_{irr}$  & I  \\
        (MeV) & & (mg/cm$^2$) &    (hrs) & (nA)  \\
\noalign{\smallskip}\hline\noalign{\smallskip}
        21.54$\pm$0.06 & &  8.30$\pm$0.04   & 1.25  & 100  \\
        19.63$\pm$0.08  & & 9.81$\pm$0.04  & 1.01  & 92    \\
        18.54$\pm$0.06  & & 7.79$\pm$0.05   & 1.05  & 99   \\ 
        17.73$\pm$0.08  & & 9.95$\pm$0.04  & 1.02  & 62   \\
        16.54$\pm$0.07  & &  8.12$\pm$0.05  & 0.82  & 111  \\
        15.71$\pm$0.09  & &  10.23$\pm$0.05 & 1.53  & 58   \\ 
        13.53$\pm$0.08  & &  8.28$\pm$0.04  & 1.93  & 62  \\
        12.57$\pm$0.08  & &  7.83$\pm$0.04  & 1.70  & 97   \\
\noalign{\smallskip}\hline
\end{tabular}
\end{table}

The incident proton flux was calculated from the charge (Q) collected  on 
the Faraday cup situated downstream of the target. To correct for beam 
fluctuations during the irradiation ($\sim28\%$), incident charge (beam current)
was recorded at regular intervals of 10 to 60 seconds.
A suitable cool-down time was allowed to ensure the permissible radioactive dose 
levels for handling the irradiated targets during counting.\\

Three HPGe detectors were used for offline measurements of $\gamma$-ray spectra of the various
irradiated samples- Detectors D1 and D2  with relative efficiency of 38\% (Bruker Baltic 
make), and detector D3 with relative efficiency of 30\% (ORTEC make) These detectors were 
calibrated using a standard $^{152}$Eu source (20105$\pm$1005 Bq). 
Irradiated targets were mounted at a distance of $\sim$10 cm from the face of
the HPGe detector and efficiency was measured in  identical geometry for all 
three detectors. The spectra were recorded using a CAEN N6724 digitizer and 
analyzed offline using the Linux Advanced MultiParameter System (LAMPS) 
software \cite{TIFR_LAMP}. \\

 \begin{figure}
    \centering
    \includegraphics[scale=0.45]{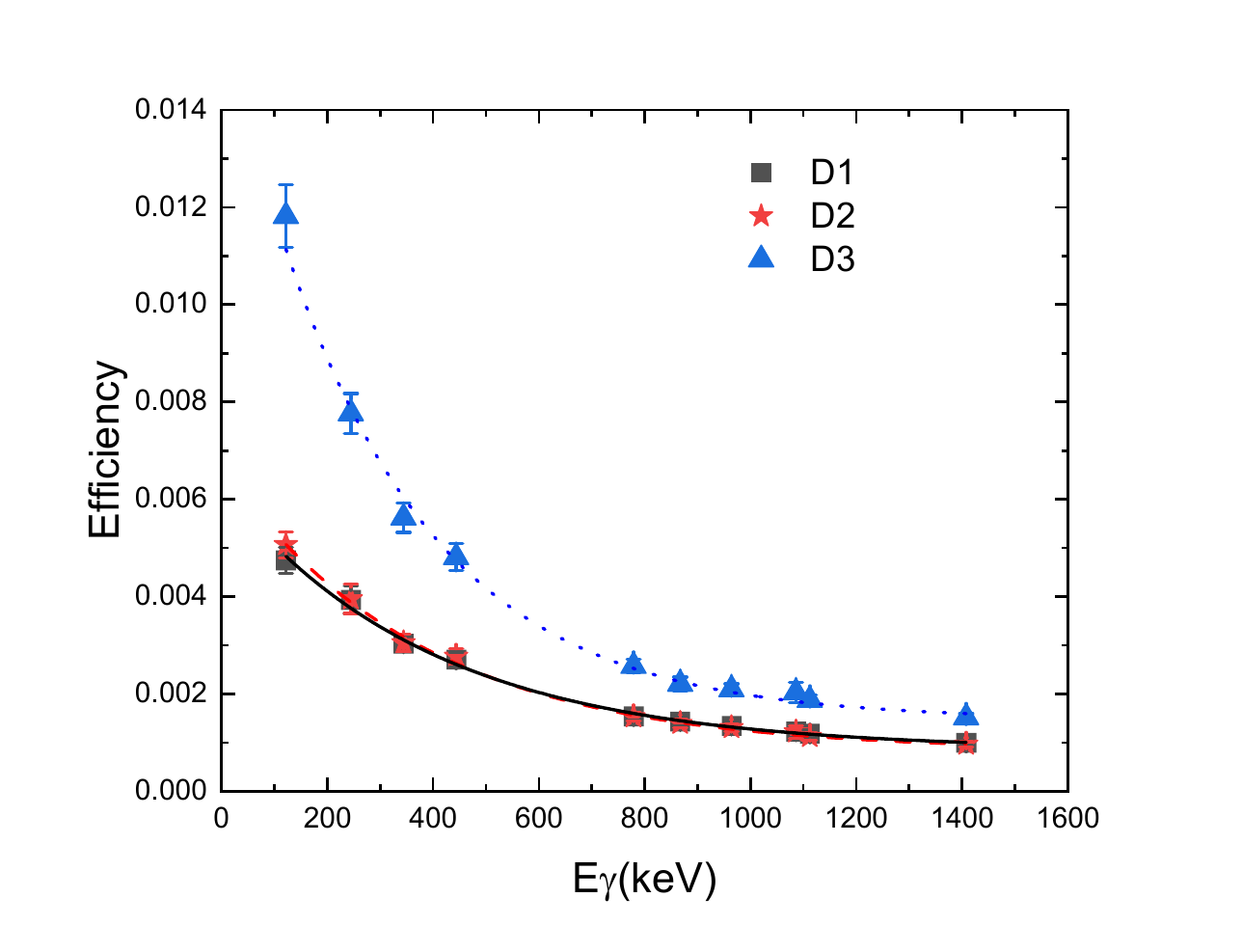}
    \caption{Absolute detector efficiency of HPGe detectors as a function of energy along with the fitted curves.}
    \label{fig:enter-label}
\end{figure}

\begin{table*}[t] 
    \centering
    \caption{Fitting parameters of efficiency curves and covariance matrix for the detectors.}
    \label{tab:fit_parameters}
    
    \begin{subtable}{\textwidth}
        \centering
        \label{tab:sub_table1}
        \begin{tabular}{cccccc} 
\toprule
Detector & Parameters & Value & \multicolumn{3}{c}{Covariance Matrix}  \\
\midrule
D1 & $\epsilon_c$  & (8.52 $\pm$ 0.44)$\times 10^{-4}$ & 1.938E-8 & & \\
   & $\epsilon_0$  & (5.42 $\pm$ 0.19)$\times 10^{-3}$ & 3.59E-9 & 3.43E-8 & \\
   & $E_0$ (keV)   & 393 $\pm$ 20 & -7.95E-4 & -2.72E-3 & 411 \\

\bottomrule
\end{tabular}
    \end{subtable}
    
    \vspace{0.5cm} 

    \begin{subtable}{\textwidth}
        \centering
        \label{tab:sub_table2}
        \begin{tabular}{cccccc} 
\toprule
& Parameters & Value &  \multicolumn{3}{c}{Covariance Matrix} \\
\midrule
D2 & $\epsilon_c$  & (8.29 $\pm$ 0.43)$\times 10^{-4}$ & 1.81E-9 & & \\
   & $\epsilon_0$  & (5.87 $\pm$ 0.21)$\times 10^{-3}$ & 3.88E-9 & 4.22E-8 & \\
   & $E_0$ (keV)  & 375 $\pm$ 19 & -6.97E-4 & -2.83E-3 & 350 \\
\bottomrule
\end{tabular}
    \end{subtable}
    
    \vspace{0.5cm} 

    \begin{subtable}{\textwidth}
        \centering
        \label{tab:sub_table3}
        \begin{tabular}{cccccc} 
\toprule
& Parameters & Value &  \multicolumn{3}{c}{Covariance Matrix} \\
\midrule
D3 & $\epsilon_c$  & (1.47 $\pm$ 0.086)$\times 10^{-3}$ & 7.38E-9 & & \\
   & $\epsilon_0$  & (1.46 $\pm$ 0.09)$\times 10^{-2}$ & 4.10E-8 & 8.13E-7 & \\
   & $E_0$ (keV)  & 297 $\pm$ 19 & -1.37E-3 & -1.45E-2 & 371 \\
\bottomrule
\end{tabular}
    \end{subtable}

\end{table*}

The absolute detector efficiency of $\epsilon(E_\gamma)$ corresponding to  
various gamma rays observed from the $^{152}Eu$ source was calculated using
the equation: 
\begin{equation}
    \epsilon(E_\gamma) =\frac{N_\gamma}{N_0 I_\gamma t_m}  \label{ Eq-eff },  
\end{equation}

\noindent where  $N_0$ is the source activity, $N_\gamma$ and $I_\gamma$ are the
photopeak counts and branching ratio of $\gamma$ ray of interest, and  $ t_m$ is
the counting time. 
The efficiency data was fitted to an exponential function: 

\begin{equation}
    \epsilon (E_\gamma(i)) = \epsilon_0 exp(-E_\gamma(i)/E_0) + \epsilon_c,  
    \label{Eff-fit}
\end{equation}
where $\epsilon_0$, $E_0$ (keV) and $\epsilon_c$ are the fitting parameters of 
the equation. 
 together with fitted functions are   shown in Figure~1 for all the three detectors (D1, D2 and D3). The fitting parameters together  with the covariance matrix are listed in 
Table \ref{tab:fit_parameters}.

\begin{table*}
\caption{Decay data taken from NNDC, together with the measured half-life from the present work for the measured channels.}
\label{tab:nndc-data}
\centering
\begin{tabular}{llllllll} 
\hline\noalign{\smallskip}
Daughter  &  Reaction & Decay Mode(\%) & Q-value  &  E$_\gamma$  & I$_\gamma$   & T$_{1/2}$ & T$_{1/2} ^{expt}$ \\
 nuclei & Channels & & (MeV) & (keV) & (\%) & (hrs) & (hrs) \\
\noalign{\smallskip}\hline\noalign{\smallskip}

$^{89g}$Zr & $^{93}$Mo(p,p$\alpha$) & EC $\beta$+ (100)  & -4.359  &  909.15$\pm$0.15 & 99.04$\pm$0.03  & 78.41$\pm$0.12 & 84$\pm$5\\
         & $^{94}$Mo(p,d$\alpha$) &  & -11.812 & & &  &  \\ 
         & $^{95}$Mo(p,nd$\alpha$) &  & -19.181 & & &  &  \\
\\
$^{93g}$Tc & $^{92}$Mo(p,$\gamma$) & EC $\beta$+ (100)  & 4.087   & 1362.94$\pm$0.07 & 66.2$\pm$0.6 &2.75$\pm$0.05 &  2.98$\pm$0.08 \\
     & $^{94}$Mo(p,2n) &  & -13.661 &  &  &   & \\
     & $^{95}$Mo(p,3n) &  & -21.031 &  &  &   & \\
      &  & &  &  &  &  &   \\
      \\
$^{93m}$Tc &  $^{92}$Mo(p,$\gamma$) &  IT(77.4) & 4.087 &  391.83$\pm$0.08 & 58.3$\pm$0.9 & 0.725$\pm$0.017 & 0.70$\pm$0.01\\
     & $^{94}$Mo(p,2n) & EC $\beta$+ (22.6) & -14.415 & &  &  &  \\
    &&&&  \\
$^{94g}$Tc & $^{94}$Mo(p,n)  & EC $\beta$+ (100) & -5.038  &   702.67$\pm$0.07 & 99.6$\pm$1.8 &  4.883$\pm$0.0167 & 4.8$\pm$0.1\\
         & $^{95}$Mo(p,2n)  &   & -12.407 & &  &  &  \\
              \\      
$^{95g}$Tc & $^{95}$Mo(p,n)  & EC $\beta$+ (100) & -2.473  &  765.789$\pm$0.009 & 93.8$\pm$0.3 &  20$\pm$0.1 &    18.0$\pm$0.3\\
        & $^{96}$Mo(p,2n)  &   &  -11.627 & &  &  &    \\
        & $^{97}$Mo(p,3n)  &  &  -18.448  & &  &  &  \\
        &&&& \\ 
$^{96g}$Tc & $^{96}$Mo(p,n)  & EC $\beta$+ (100) & -3.756  &  778.22$\pm$0.4 & 99.76$\pm$0.01 &  102.72$\pm$1.68 &  97$\pm$2\\
        & $^{97}$Mo(p,2n)  &  & -10.577  & &  & &   \\
         & $^{98}$Mo(p,3n)  &  &  -19.219 & &  &  &   \\
           && &  &  & \\
           \\ 
$^{99m}$Tc  & $^{100}$Mo(p,2n)  & IT(99.9963) & -8.574 & 140.511$\pm$0.001 & 89$\pm$4 & 6.0072$\pm$0.0009 &  6.1$\pm$0.1\\
  &  $^{100}$Mo(p,pn)$\rightarrow$$^{99 }$Mo$\rightarrow$$^{99m}$Tc  & EC $\beta$+ (0.0037) & -8.290 & &  &  &  \\
  &  $^{100}$Mo(p,d)$\rightarrow$$^{99}$Mo$\rightarrow$$^{99m}$Tc  &  &  -6.607 & &  &  &  \\
  &    &  &  & &  &  &  \\
        \\          
$^{99}$Mo & $^{100}$Mo(p,2p)$\rightarrow$$^{99}$Nb$\rightarrow$$^{99}$Mo  & $\beta$-(100)  & -11.140 &  739.5 $\pm$0.017 & 12.2$\pm$0.2 & 65.924$\pm$0.006 & 63.1$\pm$0.1 \\
        & $^{100}$Mo(p,pn)  &  & -8.290  & &  &  &  \\
        & $^{100}$Mo(p,d)  &  &  -6.607  & &  &  &  \\
        \\

\noalign{\smallskip}\hline
\end{tabular}
\end{table*}

\section{Data Analysis}

\begin{figure*}
    \centering
    \includegraphics[scale=0.4]{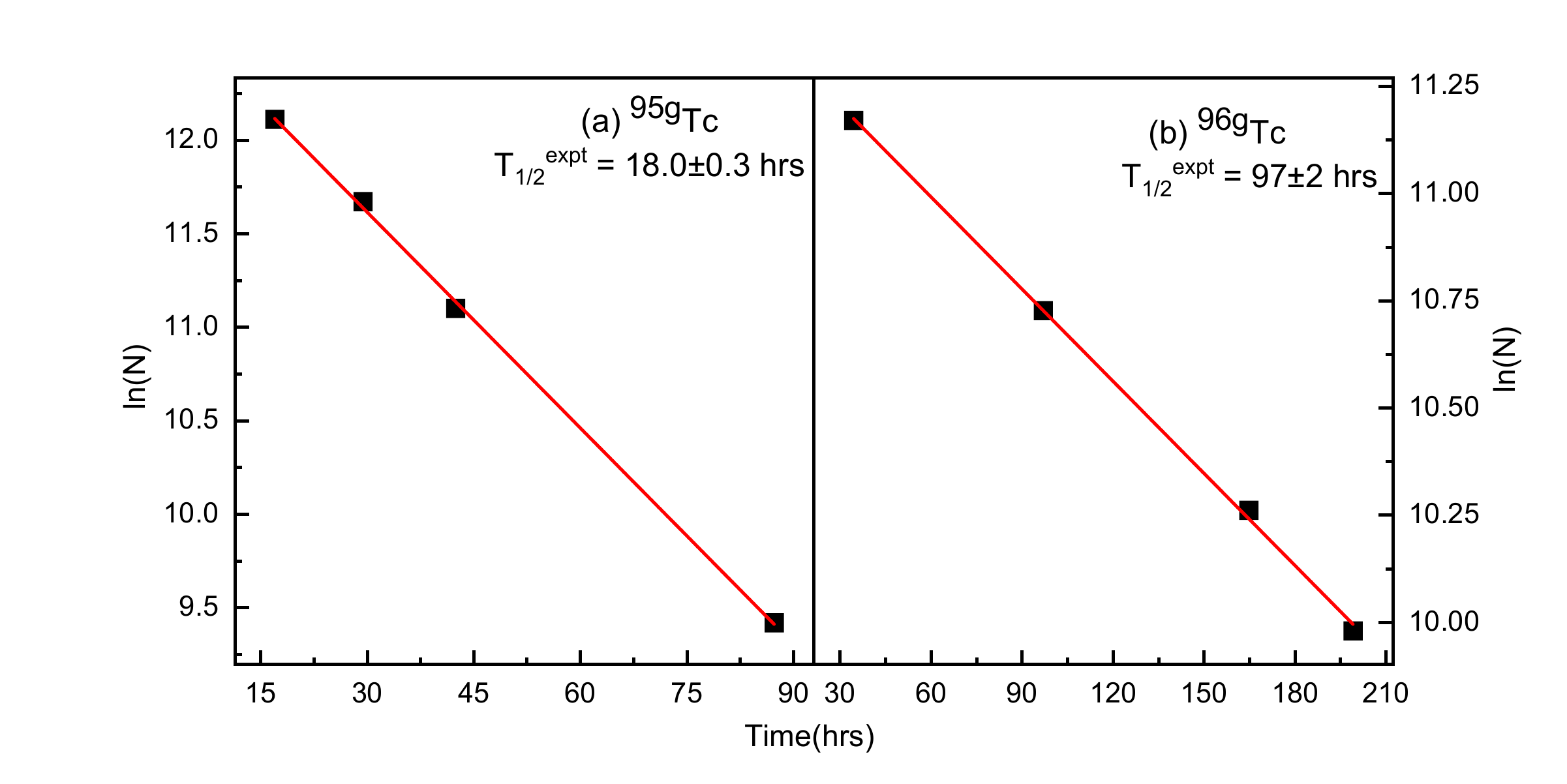}
    \caption{Measured half-life of (a)$^{95g}$Tc and (b)$^{96g}$Tc together with fit function (error bars in data points are smaller than symbol size.}
    \label{fig:halflife}
\end{figure*}

The reactions products were identified by the dominant characteristic gamma
rays. Table~3 gives a summary of the observed reaction channels and 
corresponding dominant gamma rays used for the cross section calculation. 
To ensure unique identification of reaction products, half-life was tracked by recording data at suitable
time intervals. As a typical example, the measured half-life data for $^{95g}$Tc and $^{96g}$Tc is
shown in Figure~\ref{fig:halflife}.
The measured half-life values (T$_{1/2} ^{expt}$) for all observed channels are found to be 
consistent with literature values (within measurement errors) and are also listed in Table~\ref{tab:nndc-data}.

\subsection{cross section measurement}

The production cross section of a given nuclide is calculated using the equation: \\
\begin{equation}
   \sigma=\frac{ N_\gamma\lambda}{N_t a \epsilon N_p I_\gamma T_f} \label{cs}, \\  
\end{equation}
where $N_\gamma$, I$_\gamma$, $\epsilon$ are the photopeak intensity, 
branching ratio and photopeak efficiency of the characteristic $\gamma$-ray,
respectively,  $\lambda$ is the decay constant,  $N_t$ is the number of 
target atoms  per unit area, $a$ is natural isotopic abundance of the parent 
isotope, $N_p$ is the number of protons incident per second on the target. 
The timing factor $T_f$, which is a correction  for the decay during the 
irradiation and cooldown time, is given by eqn.(\ref{tf}):
\begin{equation} 
T_f=(1-e^{-\lambda t_{irr}})e^{-\lambda t_{cool}}(1-e^{-\lambda t_{count}}), 
\label{tf} \\   
\end{equation}
where $t_{irr}$ is irradiation time, $t_{cool}$ is the cooling time and $t_{count}$ is the 
counting time. \\

The irradiation time is considered as 
$t_{irr}=\sum_1^{n} dt(i)$ such that $dt(i) \sim 10-60 {\rm sec} << T_{1/2}$. As mentioned earlier the charge incident  on the target $dq(i)$ was recorded for each interval.     To take care of  the effects of beam current fluctuations, the  cross section 
was obtained by computing the production of the radionuclide in  each interval for incident charge $dq(i)$ and correcting for the decay of the radionuclide during the time interval $dt(i)$.

Let $K ={aN_t\,\sigma}/{e}$ be the number of  nuclei produced per unit charge, where $e$ is
the electronic charge. $P(i)$, which is the number of nuclei produced in the
$i^{\text{th}}$ interval, can then be written as  

\begin{equation}
P(i) = K\,dq(i) = \frac{a N_t\,\sigma}{e} dq(i)
\label{eq:P1}
\end{equation}
Considering the  decay  of radionuclide during the same interval, the net yield Y(i)  can be expressed as \\
$$
Y(i) = Y(i-1) + P(i) - \lambda_m Y(i-1) = K \cdot X(i)
$$
The quantity X(i) is computed for each interval and the yield  at the end of irradiation is obtained
$$
Y(n) = K \cdot X(n)
$$
The cross section $\sigma$ is then extracted from   $N_{\gamma}$ and $Y(n)$ as below :

$$
Y(n) =
\frac{N_{\gamma}}
{I_{\gamma}\,\varepsilon\,e^{-\lambda t_{\mathrm{cool}}}
\left(1 - e^{-\lambda t_{\mathrm{count}}}\right)}
$$

\begin{equation}
\sigma=  \frac{N_{\gamma}}
{a N_t I_{\gamma}\,\varepsilon\,e^{-\lambda t_{\mathrm{cool}}}
\left(1 - e^{-\lambda t_{\mathrm{count}}}\right)} \frac{e}{X(n)}
\label{eq:sigma_def}
\end{equation}


\subsection{Uncertainty analysis and correlation coefficients}
The total uncertainty in the measured cross section is given by
\begin{equation}
\label{Uncertainty_formula}
\begin{split}
\left(\frac{\Delta\sigma}{\sigma}\right)^2 = &
\left(\frac{\Delta N_\gamma}{N_\gamma}\right)^2 +
\left(\frac{\Delta I_\gamma}{I_\gamma}\right)^2 +
\left(\frac{\Delta a}{a}\right)^2 \\
&+
\left(\frac{\Delta N_t}{N_t}\right)^2 +
\left(\frac{\Delta \epsilon}{\epsilon}\right)^2 +
\left(\frac{\Delta T_f}{T_f}\right)^2 .
\end{split}
\end{equation}

As mentioned earlier, targets are assumed to be uniform over the foil
area ($\sim 1 cm^2$) and the uncertainty estimated in target thickness is 
$\approx0.6\%$, which is neglected as it is much smaller compared to other factors. The uncertainty in isotopic fractions ($a$) and in  proton flux ($N_p$) are also negligibly small. 
The uncertainty in the branching
ratio ($\Delta I_\gamma$) and timing factor T$_f$ are treated as negligible whenever $\le 0.1\%$ and taken into consideration otherwise.  

Hence, the major contributions to the cross section uncertainty arise from $N_\gamma$ (photopeak 
intensity) and the detector efficiency $\epsilon$.
The efficiency $\epsilon(E_\gamma(i))$ is interpolated using  Eq.~\ref{Eff-fit}. The uncertainty $\Delta 
\epsilon(E_\gamma(i))$ in $\epsilon(E_\gamma(i))$ can be derived using the covariance matrix:
\begin {equation}
(\Delta \epsilon (E_\gamma(i)) )^2=Cov(\epsilon(E_\gamma(i)),\epsilon(E_\gamma(i)\\))
\end{equation} 
 The covariance
$\mathrm{Cov}(\epsilon(E_i), \epsilon(E_j))$ is calculated as described in Ref~\cite{otuka2017uncertainty}:

\begin{equation}
\begin{split}
    {Cov(\epsilon(E_\gamma(i)),\epsilon(E_\gamma(j)))}= e^{-\frac{E_\gamma(i) + E_\gamma(j)}{E_0}}
    (\Delta \epsilon_0)^2    + &\\ \frac{\epsilon_0^2E_\gamma(i)E_\gamma(j)}{E_0^4}e^{-
    \frac{E_\gamma(i) + E_\gamma(j)}{E_0}}(\Delta  E_0)^2  + (\Delta \epsilon_c)+ &\\ \epsilon_0 
    \frac{E_\gamma(i) + E_\gamma(j)}{E_0^2} e^{- \frac{E_\gamma(i) +  E_\gamma(j)}{E_0}} 
    Cov(\epsilon_0, E_0) + &\\ \left(e^\frac{-E_\gamma(i)}{E_0} +  e^\frac{-E_\gamma(j)}  
    {E_0}\right) Cov(\epsilon_0,\epsilon_c) + &\\ \frac{\epsilon_0}
    {E_0^2}\left(E_\gamma(i) e^\frac{-E_\gamma(i)}{E_0} + E_\gamma(j) e^\frac{-E_\gamma(j)}{E_0} 
    \right)  Cov(E_0,\epsilon_c).
\end{split}
\end{equation}
where $E_0$, $\epsilon_0$ and $\epsilon_c$ are fit parameters given in Table~\ref{tab:fit_parameters}.

The covariance between cross sections $\sigma_1$ and $\sigma_2$ at  proton energies $E_{1}$ and $E_{2}$, respectively,  is calculated using: 
\begin{equation}
    Cov(\sigma_1 , \sigma_2) = \sum_{k} \left( \Delta x_{k}^{(1)} \cdot Corr(\Delta x_{k}^{(1)}, \Delta x_{k}^{(2)}) \cdot \Delta x_{k}^{(2)} \right),
\end{equation}
\noindent
where $\Delta x_k^{(1)}$ and $\Delta x_k^{(2)}$ are the fractional uncertainties of the
$k$-th contributing term (see RHS of Eqn.~5) to the cross sections 
and $Corr(\Delta x_{k}^{(1)}, \Delta x_{k}^{(2)})$ represents the correlation between the $k$-th uncertainty components at two different proton energies.

\begin{figure*}
\centering
{\includegraphics[scale=0.45]{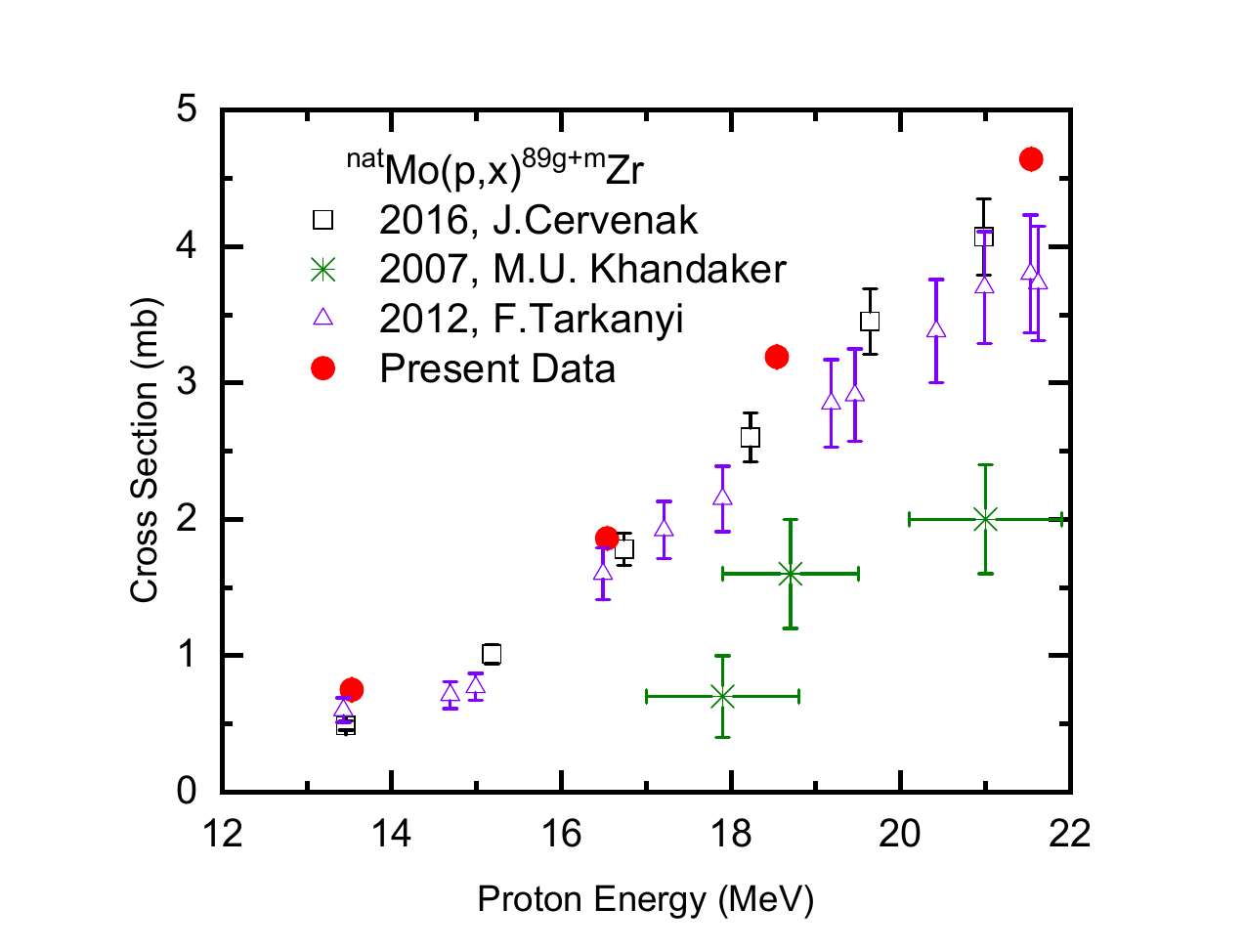}}
\hfill
  \caption{Excitation function for production of $^{89g+m}$Zr.}
  \label{fig:89g+mZr_CS}
\end{figure*}

\begin{table*} 
\centering
 \caption{cross sections for $^{nat}$Mo(p,x)$^{89g+m}$Zr reaction together with fractional uncertainty in parameters
 and correlation coefficients between $E_p$ and $\sigma$.}
    \begin{tabular}{cccccccccccc}
    \hline
    E$_p$  & $\sigma$(mb) & $\Delta \epsilon$  & $\Delta N_t $ & $\Delta N_{\gamma}$ & Total & \multicolumn{4}{c}{Correlation coefficients}\\
     (MeV) & &(\%) &(\%) &(\%) &  Uncertainty (\%) & \multicolumn{4}{c}{}\\
    \hline
       21.54 & 4.6$\pm$0.1  & 1.38    & 0.52 & 2.48 & 2.88 &1 &  &  &  &     \\
       18.54 & 3.2$\pm$0.1  & 1.38    & 0.64 & 3.67 & 3.97 & 0.17 & 1 &  &  &      \\
       16.54 & 1.9$\pm$0.2  & 1.38    & 0.59 & 8.07 & 8.21 & 0.08 &  0.06 & 1 &   &     \\
       13.53 & 0.75$\pm$0.14  & 2.35    & 0.53 & 18.93 & 19.08 & 0.06  & 0.04 & 0.02 & 1   \\
       \hline
    \end{tabular}
    \label{tab:89g+m-Zr}
\end{table*}

The uncertainties in photopeak counts and number of target atoms present in the sample for different incident proton energies are not correlated 
and hence correlation coefficient between them is zero. Since the cross section measurement at 
different proton energies involves the same characteristic gamma ray $E_{\gamma}$, the corresponding $\epsilon(E_{\gamma})$, $I_\gamma$ and  $\lambda$ are same and consequently the fractional uncertainty in $\epsilon(E_\gamma(i))$ is identical at different 
proton energies.
The correlation coefficient between cross sections at two different energies is
calculated using the formula:
\begin{equation}
    Cor(\sigma_1 , \sigma_2) = \frac{Cov(\sigma_1 , \sigma_2)}{(\Delta \sigma_1 .
    \Delta \sigma_2)} .
\end{equation}

\section{Results and Discussion}
Measured cross sections and corresponding total uncertainties for different
reaction channels 
are tabulated in Tables~\ref{tab:89g+m-Zr} to ~\ref{tab:100Mo(p,x)99mTc}. 
Fractional uncertainties of various parameters are also listed 
along with correlation coefficients. 
It should be mentioned that the accuracy of the current data is achieved through the systematic reduction of experimental uncertainties by using
thin targets, precise beam current measurements, and corrections for isomer decay contributions (wherever possible). Further, a complete covariance evaluation has been performed for the first time
for Mo(p,x) reactions, providing cross section--cross section correlation coefficients that
make the data more useful for comparisons with models and uncertainty propagation.  
To avoid the propagation of rounding off errors in the final cross section values and correlation coefficients,
the uncertainty values in individual parameters are retained to two decimal places.

The present data are compared with the 
data available in the EXFOR library. From the observed $^{nat}$Mo(p,x) cross section, $^{100}$Mo(p,x) cross section is estimated employing appropriate scaling by 
the isotopic abundance of $^{100}$Mo, for comparison with wider data 
sets. 

\begin{figure*}[htb]
    \centering
    \begin{minipage}[b]{0.44\textwidth}
        \centering
        \includegraphics[width=\linewidth]{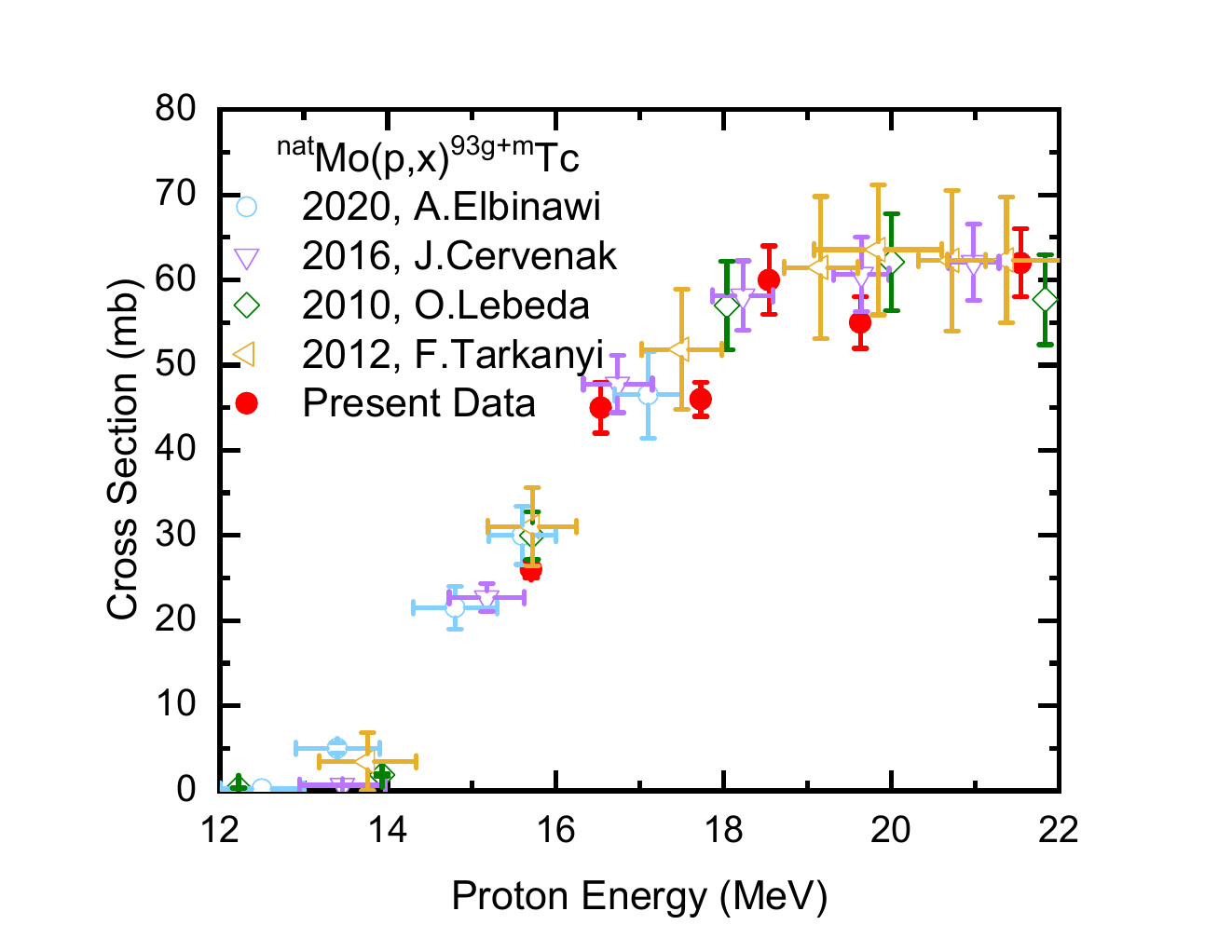}
        \caption{Excitation function of $^{\text{nat}}$Mo(p,x)$^{93\text{g+m}}$Tc reaction.}
        \label{fig:93g_m_Tc_CS}
    \end{minipage}
    \hfill
    \begin{minipage}[b]{0.51\textwidth}
        \centering
        \includegraphics[width=\linewidth]{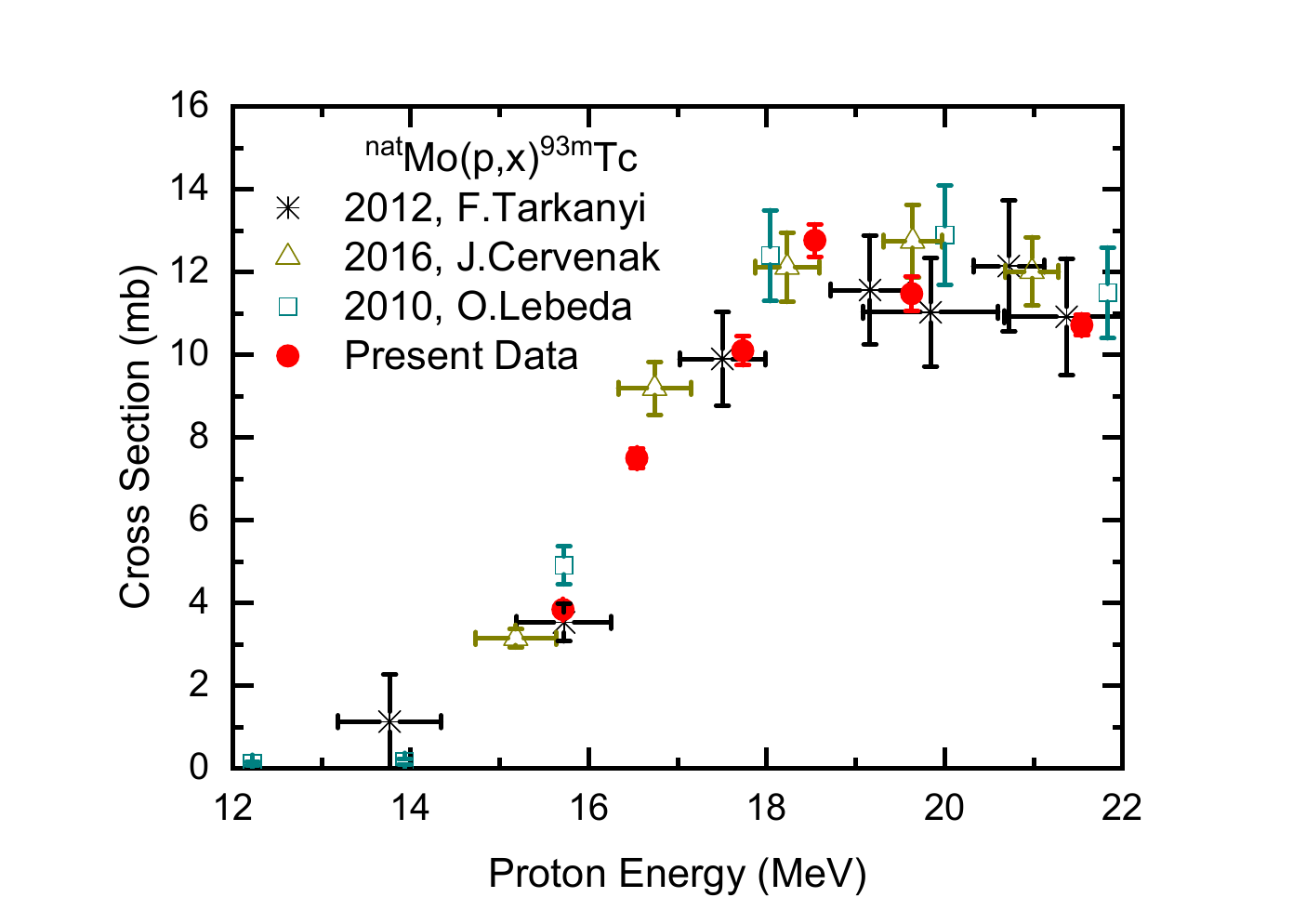}
        \caption{Excitation function of $^{\text{nat}}$Mo(p,x)$^{93\text{m}}$Tc reaction.}
        \label{fig:93m_Tc_CS}
    \end{minipage}
\end{figure*}

\subsection{Production of \(^{89g+m}\)Zr }
The $^{89g+m}$Zr is produced predominantly through $^{93}$Mo\\(p,p$\alpha$), 
$^{94}$Mo(p,d$\alpha$), and $^{95}$Mo(p,nd$\alpha$) reaction channels and the corresponding decay
data together with Q-values are summarized in Table~\ref{tab:nndc-data}. 
The $^{89}$Zr has an 
$\frac {1}{2}^-$ isomeric state at 587.8~keV with T$_{1/2}=0.08$~hrs. Due to 
relatively short half-life, the production cross section of this $^{89m}$Zr 
state could not be measured in the present experiment. This state has two 
decay branches: electron capture (EC) to $^{89}$Y, branching ratio of 
6.23$\%$ and M4 transition (IT) $^{89}Zr$ ground state ($\frac {1}{2}^+$), 
branching ratio of 93.77$\%$. 
While the latter (IT) is included in sum total of measured $^{89(g+m)}$Zr, the contribution of the former could not be assessed. 
The
measured cross sections and the corresponding correlation coefficients for various proton
energies are summarized in Table \ref{tab:89g+m-Zr}.
Figure~\ref{fig:89g+mZr_CS} compares the present data with measurements reported in Refs. ~\cite{khandaker2007measurement}, ~\cite{tarkanyi2012investigation} and ~\cite{vcervenak2016experimental}.
As can be seen in the figure,
the previous reported data are scattered with  large uncertainties (especially from~\cite{khandaker2007measurement}) and discrepancies are observed between the
measured cross sections. The large uncertainties in the incident beam energy in ~\cite{khandaker2007measurement} may be due to the use of thick targets.
On the other hand, the present
experiment used thin, high-purity molybdenum foils and proton beam with energy resolution better
than 0.1 MeV, thereby yielding more accurate excitation function. 
Present data is consistent with that of ~\cite{tarkanyi2012investigation} and ~\cite{vcervenak2016experimental},  and plays an important role in resolving  the discrepancy in previously reported data.
Additionally,  covariance analysis is carried out for the first time for this reaction, making 
these results more valuable for model evaluation and yield optimization than older datasets
without uncertainty correlations. 
The observed increasing trend of $\sigma$ with $E$ also highlights the 
importance of measurements at higher proton energies to optimize the $^{89g+m}$Zr
production. 

\begin{table*}[t]
\centering
 \caption{cross sections for $^{nat}$Mo(p,x)$^{93g+m}$Tc reaction together with fractional 
uncertainty in parameters and correlation coefficients between $E_p$ and $\sigma$.}
    \begin{tabular}{ccccccccccccccccc}
    \hline
   E$_p$  & $\sigma$(mb) & $\Delta T_f $  & $\Delta \epsilon$ & $\Delta I_{\gamma}$ & $\Delta N_t $ & 
   $\Delta N_{\gamma}$ & Total & \multicolumn{7}{c}{Correlation coefficients}\\
     (MeV) & & (\%)&(\%) &(\%) & (\%)&(\%) & Uncertainty (\%) & \multicolumn{7}{c}{}\\
    \hline
       21.54 & 53$\pm$2 & 1.92 & 2.21 & 0.91 & 0.52 & 1.57 & 3.48 &1 &  &  &  &  & &    \\
       19.63 & 42$\pm$2 & 2.06 & 2.30 & 0.91 & 0.44 & 1.40 & 3.54 & 0.39 & 1 &  &  &  &  & \\
       18.54 & 57$\pm$2 & 0.94 & 2.21 & 0.91 & 0.64 & 2.31 & 3.51 & 0.61 & 0.22 & 1 &  &  &  &    \\
       17.73 & 34$\pm$1 & 0.90 & 2.30 & 0.91 & 0.44 & 1.28 & 2.96 & 0.25& 0.76 & 0.16 & 1 &  &  & 
       \\
       16.54 & 37$\pm$2 & 1.30 & 2.21 & 0.91 & 0.59 & 3.48 & 4.46 & 0.53& 0.22 & 0.44 & 0.15 & 1 &      &   \\
       15.71 & 23$\pm$1 & 0.02  & 2.30 & 0.91 & 0.46 & 2.07 & 3.26 & 0.08& 0.53 & 0.07 & 0.64 & 0.06  & 1 &   \\
       
       \hline
    \end{tabular}
    \label{tab:natMo(p,x)93g+mTc}
\end{table*}

\begin{table*}[t]
\centering
\caption{cross sections for $^{nat}$Mo(p,x)$^{93m}$Tc reaction together with fractional 
uncertainty in parameters and correlation coefficients between $E_p$ and $\sigma$.}

    \begin{tabular}{ccccccccccccccc}
    \hline
    E$_p$  & $\sigma$(mb) & $\Delta T_f $  & $\Delta \epsilon$ & $\Delta I_{\gamma}$ & $\Delta N_t $ & $\Delta N_{\gamma}$ & Total & \multicolumn{7}{c}{Correlation coefficients}\\
     (MeV) & & (\%)&(\%) &(\%) &(\%) &(\%) & Uncertainty (\%) & \multicolumn{7}{c}{}\\
    \hline
       21.54 & 10.7$\pm$0.2 & 0.68  & 1.40  & 1.54 & 0.52 & 0.52 & 2.31 & 1 &  &  &  &  &   &   \\
       19.63 & 11.4$\pm$0.4 &  2.34 &  1.49  & 1.54 & 0.44 & 1.77 & 3.66 & 0.47 & 1 &  &  &  &      \\
       18.54 & 12.8$\pm$0.4 & 1.64 &  1.40 & 1.54 & 0.64 & 1.39 & 3.06 & 0.77 & 0.56 & 1 &  &  &     \\
       17.73 & 10.1$\pm$ 0.4  &  1.84  &  1.49 & 1.54 & 0.44 & 1.97 & 3.47 & 0.45 & 0.70 & 0.51 & 1 &  &     \\
       16.54 & 7.5$\pm$0.2 & 0.04 &  1.40  & 1.54 & 0.59 & 2.38 & 3.22 & 0.59 & 0.21 & 0.45 & 0.22 & 1 &     \\
       15.71 & 3.8$\pm$0.2 & 1.91  &  1.49 &  1.54 & 0.46 & 2.69 & 3.96 &0.40 & 0.63 & 0.46 & 0.59 & 0.19 & 1    \\
       \hline
    \end{tabular}
    \label{tab:natMo(p,x)93mTc}
\end{table*}

\begin{figure*}
\centering
{\includegraphics[scale=0.45]{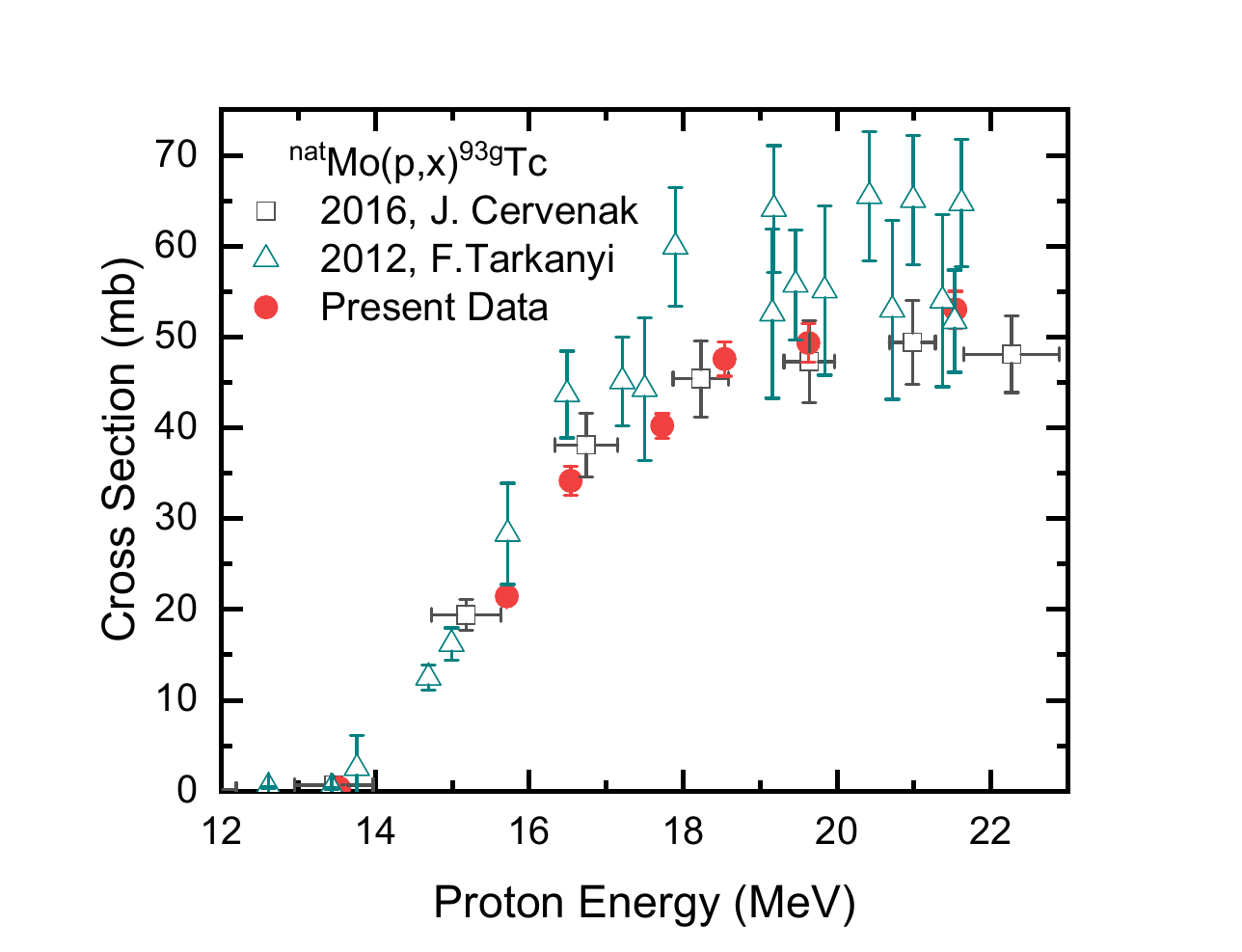}}
\hfill
  \caption{Excitation function for production of $^{93g}$Tc.}
  \label{fig:93g(m-)Tc_CS}
\end{figure*}

\begin{table*}[t]
\centering
 \caption{cross sections for $^{nat}$Mo(p,x)$^{93g}$Tc reaction together with fractional 
uncertainty in parameters and correlation coefficients between $E_p$ and $\sigma$.}
    \begin{tabular}{ccccccccccccccccc}
    \hline
   E$_p$  & $\sigma$(mb) & $\Delta T_f $  & $\Delta \epsilon$ & $\Delta I_{\gamma}$ & $\Delta N_t $ & $\Delta N_{\gamma}$ & Total & \multicolumn{7}{c}{Correlation coefficients}\\
     (MeV) & & (\%)&(\%) &(\%) & (\%)&(\%) & Uncertainty (\%) & \multicolumn{7}{c}{}\\
    \hline
       21.54 & 53$\pm$2 & 1.92 & 2.21 & 0.91 & 0.52 & 2.13 & 3.76 &1 &  &  &  &  & &    \\
       19.63 & 49$\pm$2 & 2.06 & 2.30 & 0.91 & 0.44 & 2.82 & 4.30 & 0.30 & 1 &  &  &  &  & \\
       18.54 & 48$\pm$2 & 0.94 & 2.21 & 0.91 & 0.64 & 2.93 & 3.94 & 0.51 & 0.16 & 1 &  &  &  &    \\
       17.73 & 40$\pm$1 & 0.90 & 2.30 & 0.91 & 0.44 & 2.02 & 3.35 & 0.20& 0.55 & 0.13 & 1 &  &  &   \\
       16.54 & 34$\pm$2 & 1.30 & 2.21 & 0.91 & 0.59 & 3.80 & 4.71 & 0.46& 0.17 & 0.37 & 0.13 & 1 &  &   \\
       15.71 & 21$\pm$1 & 0.02  & 2.30 & 0.91 & 0.46 & 1.78  & 3.08 & 0.07& 0.46 & 0.07 & 0.59 & 0.06 & 1 &   \\
       13.53 & 0.34$\pm$0.06 &0.47& 2.30 & 0.91 & 0.53 &17.07& 17.27& 0.03& 0.10 & 0.02 & 0.11 & 
       0.02 & 0.12 & 1 \\
       \hline
    \end{tabular}
    \label{tab:natMo(p,x)93gTc(m-)}
\end{table*}

\subsection{Production of \(^{93}\)Tc} 

The $^{93}$Tc can be produced either in the ground state or in the isomeric state via $^{92}$Mo(p,$\gamma$), $^{94}$Mo(p,2n)
and $^{95}$Mo(p,3n) reactions as listed in Table~\ref{tab:nndc-data}. However, $^{92}$Mo (p,$\gamma$) cross sections are very
small and due to high Q value contribution from $^{95}$Mo(p,3n) is relevant ($\sim10\%$) only at the highest incident energy. 
The isomeric state $^{93\text{m}}$Tc ($1/2^{-}$, $T_{1/2}=0.725$ hrs) de-excites either via an M4 gamma transition to $^{93\text{g}}$Tc
(f$\sim$77.4\%) or via electron capture to $^{93}$Mo ($\sim$22.6\%). Given the comparable half-lifes of $^{93\text{m}}$Tc and $^{93\text{g}}$Tc,
the measured yield of 1362~keV gamma ray (from the decay of $^{93\text{g}}$Tc) has contributions from the directly produced ground state
 ($^{93\text{g}}$Tc) and the isomeric decay $^{93\text{m}}$Tc.
 Most previous studies have reported combined $^{93\text{g+m}}$Tc yield, making it
 difficult to assess the independent ground state production in \(^{93g}\)Tc.  In a couple of cases e.g.
~\cite{tarkanyi2012investigation},~\cite{vcervenak2016experimental} the \(^{93g}\)Tc cross section is estimated either by
 neglecting or approximating decay corrections. In the present work, the cross section for the  direct production  of
 $^{93\mathrm g}$Tc is extracted by proper subtraction of the  \(^{93m}\)Tc isomer decay contribution  from the cumulative
 yield \(^{93g+m}\)Tc.

\subsubsection{Production of \(^{93g+m}\)Tc} 
The $N_{1362}$,  total observed photopeak counts for 1362~keV gamma ray during counting, can be expressed as 

\begin{equation}
N_{1362} = \varepsilon\, I_{\gamma}
\int_{t_1}^{t_2}
\frac{d}{dt}
\left[
Y_{g}^{\mathrm{direct}}(t) + Y_{m \rightarrow g}(t)
\right] \, dt
\label{eq:N1362_counts}
\end{equation}
where 
 $Y_{g}^{\mathrm{direct}}$ and $Y_{m\rightarrow g}$ represent the yield of $^{93\mathrm{g}}$Tc produced directly and from the decay of $^{93\mathrm{m}}$Tc, respectively.
The total \(^{93g+m}\)Tc cross section values obtained from net yield of 1362 keV gamma ray are given in Table \ref{tab:natMo(p,x)93g+mTc} together
with the corresponding correlation coefficients for different incident
proton energies. The data are plotted in Fig.~\ref{fig:93g_m_Tc_CS}. It can be seen from the figure that the cross sections for
the $^{nat}$Mo(p,x)$^{93\text{g+m}}$Tc reaction,  increase  from about 12 to 18~MeV and are nearly constant thereafter.
The present data  is in  good agreement
with the previously reported measurements in Refs.~\cite{tarkanyi2012investigation}-\cite{ahmed2022study}.
It should be emphasized that the reduced uncertainties in the present data provide a 
more reliable evaluation of the $^{93\text{g+m}}$Tc production channel.

\begin{figure*}
\centering
{\includegraphics[scale=0.45]{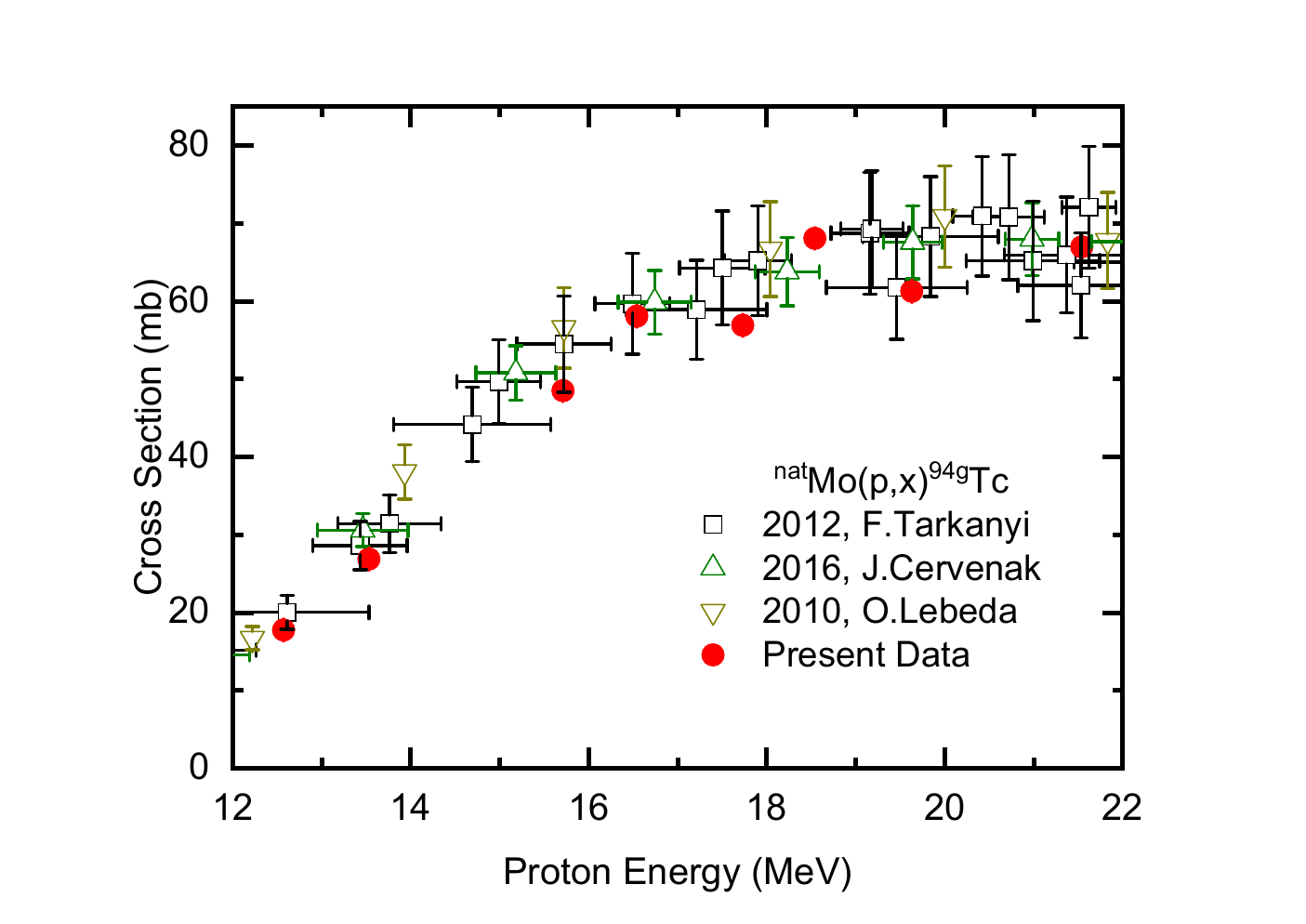}}
\hfill
  \caption{Excitation function for production of $^{94g}$Tc.}
  \label{fig:94gTc_CS}
\end{figure*}

\begin{table*}[t]
\centering
\caption{cross sections for $^{nat}$Mo(p,x)$^{94g}$Tc reaction together with fractional 
uncertainty in parameters and correlation coefficients between $E_p$ and $\sigma$.}
    \begin{tabular}{cccccccccccccccc}
    \hline
     E$_p$  & $\sigma$(mb) & $\Delta T_f $  & $\Delta \epsilon$ & $\Delta I_{\gamma}$ & $\Delta N_t $ & $\Delta N_{\gamma}$ & Total & \multicolumn{8}{c}{Correlation coefficients}\\
     (MeV) & &(\%) &(\%) &(\%) &(\%) &(\%) & Uncertainty (\%) & \multicolumn{8}{c}{}\\
    \hline
21.54 & 67.0$\pm$0.3 &  0.27 & 1.64 & 1.81 & 0.52 & 3.24 & 4.09 & 1 &  &  &  &  &  &  &  \\
19.63 & 61.3$\pm$0.2 & 0.23  & 1.73 & 1.81 & 0.44 & 1.22 & 2.82 & 0.29 & 1 &  &  &  &  & &    \\
18.54 & 68.1$\pm$0.2 & 0.25  & 1.64 & 1.81 & 0.64 & 2.22 & 3.37 &0.44& 0.35 & 1 &  &  &  &  &   \\
17.73 & 56.9$\pm$0.2 &  0.24 & 1.73  & 1.81 & 0.44 & 0.93 & 2.72 &0.30 & 0.82 & 0.36 & 1 &  &  &  &   \\
16.54 & 58.1$\pm$0.2 & 0.29  & 1.64 & 1.81 & 0.59 & 2.89 & 3.84 &0.38 & 0.31 & 0.47 & 0.32 & 1 &  &  &   \\
15.71 & 48.5$\pm$0.1 &  0.14 & 1.73 & 1.81 & 0.46 & 0.83 & 2.68 &0.30 & 0.83 & 0.37 & 0.86 & 0.32 & 1 &  &  \\
13.53 & 26.9$\pm$0.1 &  0.10 & 1.73 & 1.81 & 0.53 & 1.18 & 2.82&0.29 & 0.79 & 0.35 & 0.82 & 0.30 & 0.83 & 1 &   \\
12.57 & 17.8$\pm$0.1 & 0.25  & 1.64  & 1.81 & 0.55 & 2.26 & 3.38 & 0.43& 0.35 & 0.53 & 0.36 & 0.46 & 0.36 & 0.35 & 1  \\
       \hline
    \end{tabular}
    \label{tab:94gTc}
\end{table*}

\subsubsection{Production of \(^{93m}\)Tc:}
The $^{93\text{m}}$Tc isomer is important for medical isotope studies because its decay directly
contributes to the production of $^{93\text{g}}$Tc. The measured cross section values along with the correlation coefficients 
for various incident proton energies are listed in Table~\ref{tab:natMo(p,x)93mTc}.
The  comparison of the present data with the earlier data reported in Refs.~\cite{tarkanyi2012investigation}-\cite{lebeda2010new}
is shown in Figure ~\ref{fig:93m_Tc_CS}. 
Present measurements show overall agreement with the 
reported data, while a significant improvement in the in data quality is evident since energy uncertainties are considerably reduced with use of thin targets. 


\begin{figure*}[htb]
    \centering
    \begin{minipage}[b]{0.49\textwidth}
        \centering
        \includegraphics[width=\linewidth]{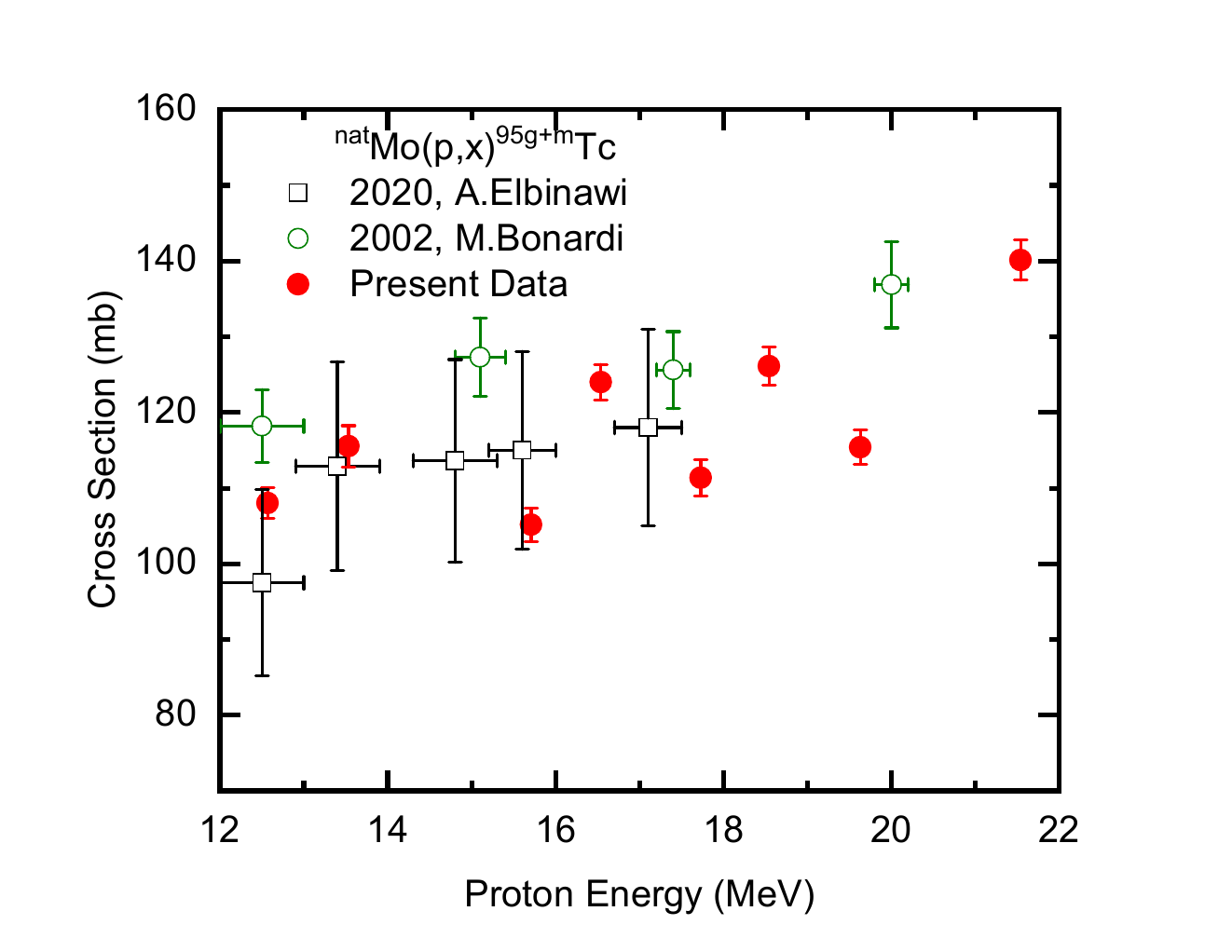}
        \caption{Excitation function for production of $^{95g+m}$Tc.}
        \label{fig:95g+mTc_CS}
    \end{minipage}
    \hfill
    \begin{minipage}[b]{0.5\textwidth}
        \centering
        \includegraphics[width=\linewidth]{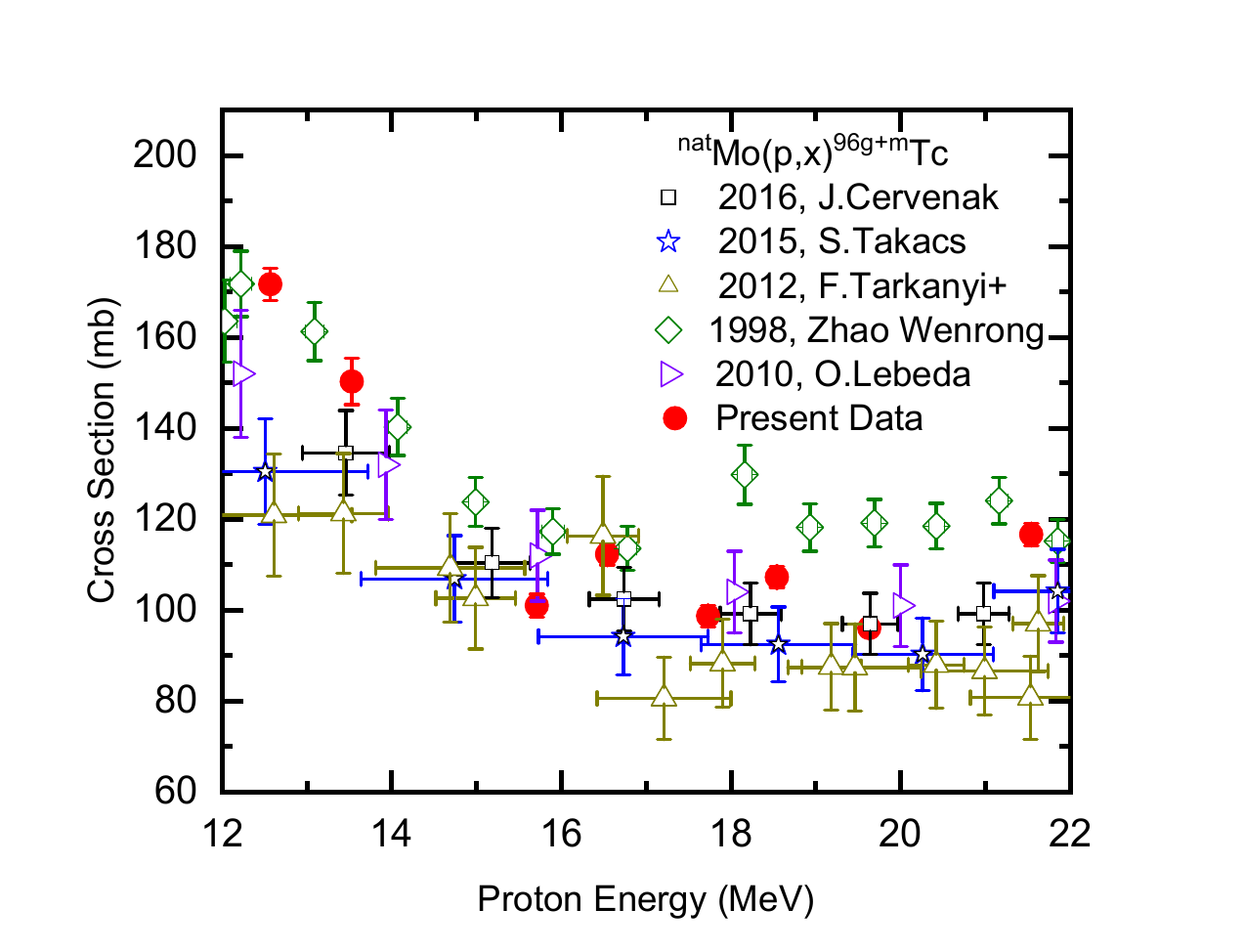}
        \caption{Excitation function for production of $^{96g+m}$Tc.}
        \label{fig:96g+mTc_CS}
    \end{minipage}
\end{figure*}

 \begin{table*}[t]
\centering
\caption{cross sections for $^{nat}$Mo(p,x)$^{95g+m}$Tc reaction together with fractional 
uncertainty in parameters and correlation coefficients between $E_p$ and $\sigma$.}
    \begin{tabular}{cccccccccccccccc}
    \hline
   E$_p$  & $\sigma$ & $\Delta T_f $  & $\Delta \epsilon$ & $\Delta I_{\gamma}$ & $\Delta N_t $ & $\Delta N_{\gamma}$ & Total & \multicolumn{8}{c}{Correlation coefficients}\\
     (MeV) & (mb)&(\%) &(\%) &(\%) &(\%) &(\%) & Uncertainty (\%) & \multicolumn{8}{c}{}\\
    \hline
21.54 & 140$\pm$3 & 0.20 & 1.58 & 0.32 & 0.52 & 0.85 & 1.91 & 1 &  &  &  &  &  &  &  \\
19.63 & 115$\pm$3 & 0.10 & 1.68 & 0.32 & 0.44 & 0.84 & 1.95 &0.03 & 1 &  &  &  &  & &    \\
18.54 & 126$\pm$3 & 0.18 & 1.58 & 0.32 & 0.64 & 0.93 & 1.98 & 0.70 & 0.03 & 1 &  &  &  &  &   \\
17.73 & 111$\pm$2 & 0.19 & 1.54 & 0.32 & 0.44 & 1.40 & 2.16 & 0.03 & 0.64 & 0.03 & 1 &  &  &  &   \\
16.54 & 124$\pm$2  & 0.18 & 1.58 & 0.32 & 0.59 & 0.79 & 1.90 & 0.73 & 0.03 & 0.70 & 0.03 & 1 &  &  &   \\
15.71 & 105$\pm$2  & 0.23 & 1.68 & 0.32 & 0.46 & 1.10 & 2.09 & 0.04 & 0.72 & 0.03 & 0.60 & 0.04 & 1 &  &   \\
13.53 & 116$\pm$3  & 0.23 & 1.54 & 0.32 & 0.53 & 1.65 & 2.36 & 0.03 & 0.59 & 0.03 & 0.50 & 0.03 & 0.56 & 1 & \\
12.57 & 108$\pm$2  & 0.24 & 1.58 & 0.32 & 0.55 & 0.75 & 1.88 & 0.74 & 0.03 & 0.71 & 0.04 & 0.74 & 0.04 & 0.04 & 1  \\
       \hline
    \end{tabular}
    \label{tab:95g+mTc}
\end{table*}

\subsubsection{Production of \(^{93g}\)Tc} 
The determination of the direct 
$^{93\text{g}}$Tc production cross section represents one of the key strengths of the present work. 
The second term $Y_{m\rightarrow g}$ in equation [\ref {eq:N1362_counts}]
is computed using the $\sigma(^{93m}$Tc) obtained in
subsection 4.2.2. Employing the procedure described in section 3.1, 
yield of $^{93m}$Tc at the end of the $i^{th}$ interval, $Y_{93m}(i)-$,  is given by 

\begin{equation}
Y_{93m}(i) = Y_{93m}(i-1) + K_m dq(i) - \lambda_m Y_{93m}(i-1) dt(i)
\label{93mTc_yield}
\end{equation}
where $K_m$ is the production rate of $^{93m}$Tc per unit charge and $\lambda_m$ is the decay
constant.  Given the fraction $f$ feeding the ground state from the isomer decay and  the decay
constant for the ground state $\lambda_g$, the yield of the $^{93g}$Tc from the isomer decay at the
end of the $i^{th}$ interval can be expressed as 

\begin{equation}
\begin{split}
Y_{m\rightarrow g}(i)
=\;&
Y_{m\rightarrow g}(i-1)
+
f\,\lambda_{m}\,Y_{93\mathrm{m}}(i-1)\,dt(i)
\\
&-
\lambda_{g}\,Y_{m\rightarrow g}(i-1)\,dt(i),
\end{split}
\label{eq:Ym_to_g_recursion}
\end{equation}
 Thus, $Y_{m\rightarrow g}$ can be computed at the end of the irradiation. Further, the $Y_{m\rightarrow g}$ is computed during the cooling time and the counting time following the same procedure, by setting the production rate of $^{93m}Tc$ to zero.  
The counting period $t_{count}$ is divided into $J_{cnt}$ intervals of duration $dt$ each and  total number of
ground-state decays ($D_g$) are computed as
\begin{equation}
D_{g} = \sum_{j=1}^{J_{\mathrm{CNT}}}
\lambda_{g}\,
Y_{m \rightarrow g}(j-1)\,dt(j),
\label{eq:Dg_sum}
\end{equation}
The contribution from   $m\rightarrow g$ feeding to the photopeak yield of 1362 keV is  given by 

\begin{equation}
N_{1362}(m \rightarrow g)
=
\varepsilon\, I_{\gamma}\, D_{g},
\label{eq:N1362_m_to_g}
\end{equation}
The direct ground–state contribution to 1362 photopeak counts is  obtained as  
\begin{equation}
N_{1362}^{\mathrm{direct}}
=
N_{1362}
-
N_{1362}(m \rightarrow g),
\label{eq:N1362_direct}
\end{equation}
Again employing the same procedure described in section 3.1, the direct production cross section of $^{93g}$Tc, $\sigma_g$, is extracted.
This step-by-step calculation of production and decay of both $^{93m}$Tc and $^{93g}$Tc throughout irradiation, cooling, and counting,  properly accounts for beam fluctuations and decay corrections, yielding accurate values of $\sigma(^{93g}$Tc).

The extracted $\sigma (^{93g}$Tc)  values along with their correlation coefficients at different proton energies are listed in Table 
\ref{tab:natMo(p,x)93gTc(m-)}.
and plotted in Figure
\ref{fig:93g(m-)Tc_CS}. 
As is evident from the Figure
the cross section increases rapidly upto 18~MeV and is nearly constant at higher energies. 
It should be mentioned that the lowest beam energy of 13.53 MeV in the present experiment is significantly lower than the Q-
value of the $^{94}$Mo (p,2n) $^{93m}$Tc reaction and hence the observed cross section at this energy corresponds only to the
production of $^{93g}$Tc. 
The data of ~\cite{tarkanyi2012investigation} and ~\cite{vcervenak2016experimental} are  also shown for a comparison,
although in these works the decay corrections are not properly incorporated. 
Measurements at higher energies are desirable to understand the saturation value of $^{93g}$Tc production.

\begin{table*}
\centering
\caption{cross sections for $^{nat}$Mo(p,x)$^{96g+m}$Tc reaction together with fractional 
uncertainty in parameters and correlation coefficients between $E_p$ and $\sigma$.}
    \begin{tabular}{cccccccccccccccc}
    \hline
    E$_p$  & $\sigma$ & $\Delta T_f $  & $\Delta \epsilon$   & $\Delta N_ t$ & $\Delta N_{\gamma}$ & Total & \multicolumn{8}{c}{Correlation coefficients}\\
     (MeV) & (mb)&(\%) &(\%)  &(\%) &(\%) & Uncertainty (\%) & \multicolumn{8}{c}{}\\
    \hline
21.54 &117$\pm$3 & 1.13 & 1.57 & 0.52 & 0.74 & 2.13 & 1 &  &  &  &  &  &  &  \\
19.63 &96$\pm$2 & 0.58 &  1.66 & 0.44 & 0.85 & 2.00 & 0.15 & 1 &  &  &  &  & &    \\
18.54 &107$\pm$2 & 1.04 & 1.57 & 0.64 & 0.81 & 2.15 & 0.79 & 0.14 & 1 &  &  &  &  &   \\
17.73 &99$\pm$2 & 1.06 & 1.53 & 0.44 & 1.44  & 2.40 & 0.23 & 0.66 & 0.21 & 1 &  &  &  &   \\
16.54 &112$\pm$2 & 0.94 & 1.57 & 0.59 & 0.82 & 2.09 & 0.79 & 0.13 & 0.77 & 0.20 & 1 &  &  &   \\
15.71 &101$\pm$3 & 1.28 & 1.66 & 0.46 & 1.28 & 2.50 & 0.27 & 0.70 & 0.25 & 0.65 & 0.23 & 1 &  &   \\
13.53 &150$\pm$5 & 0.92 & 2.74 & 0.53 & 1.71 & 3.40 & 0.14 & 0.75 & 0.13 & 0.63 & 0.12 & 0.67 & 1 &   \\
12.57 &172$\pm$4 & 0.88 & 1.57 & 0.55 & 0.85 & 2.06 & 0.78 & 0.12 & 0.76 & 0.19 & 0.76 & 0.22 & 0.12 & 1  \\
       \hline
    \end{tabular}
    \label{tab:96g+mTc}
\end{table*}

\begin{figure*}[ht!]
\centering
\includegraphics[width=0.95\textwidth]{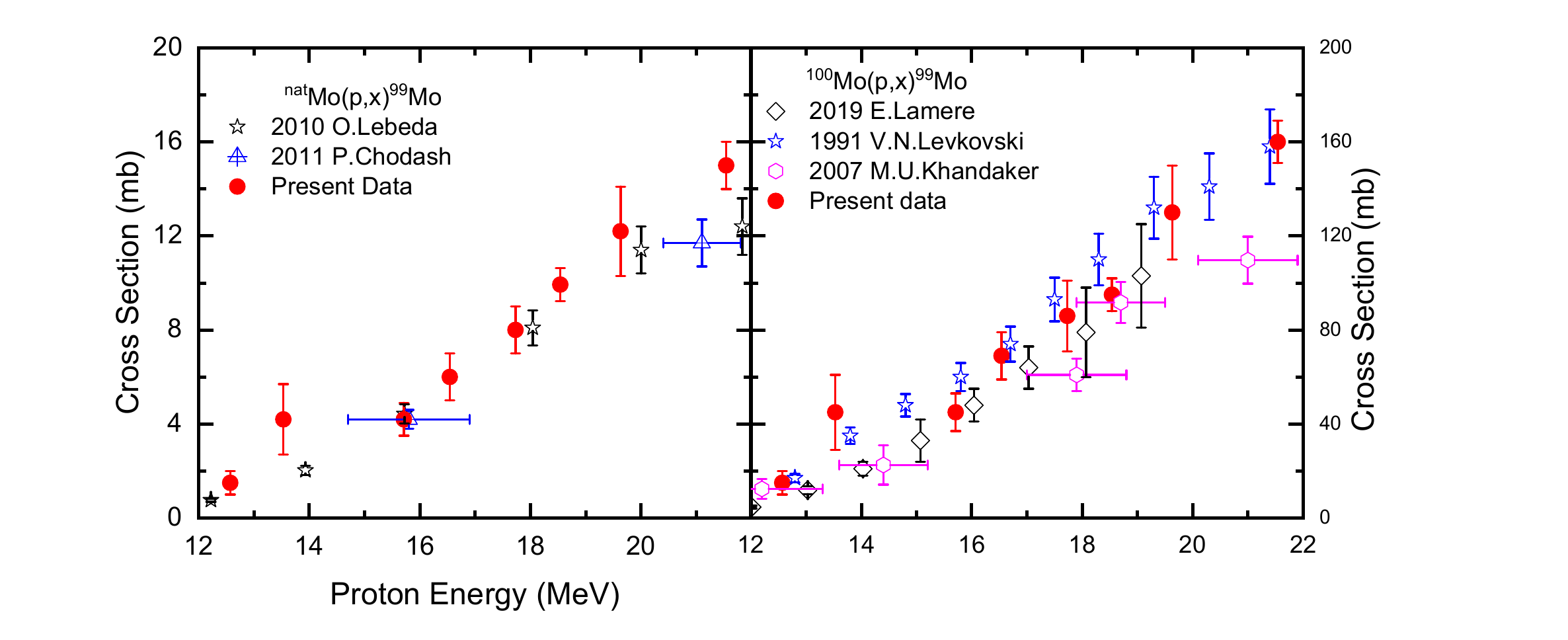}
\vspace{0.5cm} 
\caption{Excitation function for (a)$^{nat}$Mo(p,x)$^{99}$Mo and (b) $^{100}$Mo(p,x)$^{99}$Mo reactions.}
\label{fig:99Mo_CS}
\vspace{0.5cm} 
 \end{figure*}

\subsection{Production of \(^{94g}\)Tc:} 
The $^{94g}$Tc production occurs  predominantly through the 
$^{95}$Mo(p,2n) and $^{94}$Mo(p,n) channels as listed in 
Table~\ref{tab:nndc-data}. The measured cross section values of $^{94g}$Tc production are listed in
Table~\ref{tab:94gTc}, together with the corresponding correlation coefficients.
Figure \ref{fig:94gTc_CS} shows a comparison of the present  excitation function with previously reported
measurements. Measurements reported in Refs.~\cite{elbinawi2019study,ahmed2022study,alharbi2011activation,anees2019reexamination,bonardi2002thin} show a 
similar general trend, but have large uncertainties as
well as large spread and hence have not been included in the comparison.
The present data shows an overall agreement with the excitation function as that of 
~\cite{tarkanyi2012investigation,vcervenak2016experimental,lebeda2010new}, with much smaller
uncertainties. 

\subsection{Production of \(^{95g+m}\)Tc:} 
In  the $^{nat}$Mo(p,x) reactions, $^{95}$Tc can be produced either in the ground state or in the isomeric state $^{95\text{m}}$Tc ($1/2^{-}$). The isomeric state decays to the ground state 
either by a gamma transition $1/2^{-}\rightarrow 9/2^{+}$ (branching ratio of 3.88\%), or  by electron capture to $^{95}$Mo (branching ratio 96.12\%).
Due to the long half-life of $^{95\text{m}}$Tc (T$_{1/2}$ = 1464 hrs), its
production cross section could not be measured directly. However, it contributes to the measured yield of 765.8 keV gamma ray, observed in the decay of  $^{95\text{g}}$Tc. Hence, the present data represent the combined $^{95(g+m)}$Tc production.
Figure \ref{fig:95g+mTc_CS} compares the measured excitation function  of 
$^{nat}$Mo(p,x)$^{95\text{g+m}}$Tc  with previously reported data. It should be
pointed out that in the energy window of interest (12-22 MeV), measurements of 
Bonardi et
al.~\cite{bonardi2002thin} have very sparse data points, while more recent data 
~\cite{elbinawi2019study}  are limited to $E_p<$ 17~MeV and have large uncertainties
(both in $E$ and $\sigma$).  The present data indicates  that cross section is 
nearly constant upto 20 MeV. Further measurements are needed at higher energy as 
about 20\% increase is observed at the highest energy. 
The cross section values along with respective correlation coefficients are summarized in
Table~\ref{tab:95g+mTc}.

\begin{table*}[t]
\centering
\caption{cross sections for $^{nat}$Mo(p,x)$^{99}$Mo reaction together with fractional 
uncertainty in parameters and correlation coefficients between $E_p$ and $\sigma$.}
    \begin{tabular}{cccccccccccccccc}
    \hline
     E$_p$  & $\sigma$  & $\Delta \epsilon$ & $\Delta I_{\gamma}$  & $\Delta N_t $ & $\Delta N_{\gamma}$ & Total & \multicolumn{8}{c}{Correlation coefficients}\\
     (MeV) & (mb)&(\%) &(\%) &(\%) &(\%) & Uncertainty (\%) & \multicolumn{8}{c}{}\\
    \hline
21.54 & 15$\pm$1 & 1.61 & 1.64 & 0.52 & 5.10 & 5.62 &1 &  &  &  &  &  &  &  \\
19.63 & 12.2$\pm$1.9 & 1.70 & 1.64 & 0.44 & 15.32 & 15.50& 0.03 & 1 &  &  &  &  & &    \\
18.54 & 9.3$\pm$0.7 & 1.61 & 1.64 & 0.64 & 7.07  & 7.46 & 0.13 & 0.02 & 1 &  &  &  &  &   \\
17.73 & 8$\pm$1  & 1.57 & 1.64 & 0.44 & 17.77& 17.92& 0.03 & 0.02 & 0.02 & 1 &  &  &  &   \\
16.54 & 6$\pm$1  & 1.61 & 1.64 & 0.59 & 14.53& 14.72& 0.06 & 0.01 & 0.05 & 0.01 & 1 &  &  &   \\
15.71 & 4.2$\pm$0.7  & 1.70 & 1.64 & 0.46 & 16.55& 16.73& 0.03 & 0.02 & 0.02 & 0.02 & 0.01 & 1 &  &   \\
13.53 & 4.2$\pm$1.5  & 2.84 & 1.64 & 0.53 & 36.19& 36.35& 0.01 & 0.01 & 0.01 & 0.01 & 0.01 & 0.01 & 1 &   \\
12.57 & 1.5$\pm$ 0.5 & 1.61 & 1.64 & 0.55 & 32.84 & 32.92 & 0.03 & 0.01 & 0.02 & 0.005 & 0.01 & 0.005 & 0.002 & 1  \\
       \hline
    \end{tabular}
    \label{tab:natMo(p,x)99Mo}
\end{table*}

\begin{table*}[t]
\centering
\caption{cross sections for $^{100}$Mo(p,x)$^{99}$Mo reaction together with fractional 
uncertainty in parameters and correlation coefficients between $E_p$ and $\sigma$.}
    \begin{tabular}{cccccccccccccccc}
    \hline
    E$_p$  & $\sigma$  & $\Delta \epsilon$ & $\Delta I_{\gamma}$ & $\Delta a$ & $\Delta N_t $ & $\Delta N_{\gamma}$ & Total & \multicolumn{8}{c}{Correlation coefficients}\\
     (MeV) & (mb)&(\%) &(\%) & (\%) &(\%) &(\%) & Uncertainty (\%) & \multicolumn{8}{c}{}\\
    \hline
21.54 &160$\pm$9 & 1.61 & 1.64 & 0.67 & 0.52 & 5.10  & 5.66  & 1 &  &  &  &  &  &  &  \\
19.63 &130$\pm$20 & 1.70 & 1.64 & 0.67 & 0.44 & 15.32& 15.52 & 0.04 & 1 &  &  &  &  &  &    \\
18.54 &95$\pm$7  & 1.61 & 1.64 & 0.67 & 0.64 & 7.07  & 7.49  & 0.14 & 0.03 & 1 &  &  &  &  &   \\
17.73 &86$\pm$15 & 1.57 & 1.64 & 0.67 & 0.44 & 17.77& 17.93 & 0.03 & 0.02 & 0.02 & 1 &  &  &  &   \\
16.54 &69$\pm$10 & 1.61 & 1.64 & 0.67 & 0.59 & 14.53& 14.73 & 0.07 & 0.01 & 0.05 & 0.01 & 1 &  &  &   \\
15.71 &45$\pm$8  & 1.70 & 1.64 & 0.67 & 0.46 & 16.55 & 16.74 & 0.03 & 0.02 & 0.02 & 0.02 & 0.01 & 1 &  &   \\
13.53 &45$\pm$16 & 2.84 & 1.64 & 0.67 & 0.53 & 36.19 & 36.35 & 0.02 & 0.01 & 0.01 & 0.01 & 0.01 & 0.01 & 1 & \\
12.57 &15$\pm$5  & 1.61 & 1.64 & 0.67 & 0.55 & 32.84 & 32.93& 0.03 & 0.01 & 0.02 & 0.01 & 0.01 & 0.01 & 0.003 & 1  \\
       \hline
    \end{tabular}
    \label{tab:100Mo(p,x)99Mo}
\end{table*}

\subsection{Production of \(^{96g+m}\)Tc:} 
The production of $^{96}$Tc occurs primarily through the
$^{96}$Mo(p,n), $^{97}$Mo(p,2n), and $^{98}$Mo(p,3n) reaction channels, as listed in 
Table~\ref{tab:nndc-data}. 
In the case of $^{96}$Tc also, the measured ground state data 
has contribution from the 
isomeric state $^{96\text{m}}$Tc ($4^{+}$, T$_{1/2}$ = 0.86 hrs), which could not be extracted separately.
The isomer decays predominantly via an M3 gamma transition to the ground state (
IT $98\%$), while a small fraction ($2\%$) decays to $^{96}$Mo following an EC.
Consequently, the measured cross sections $^{96}$Tc represent the combined 
production of $^{96(g+m)}$Tc. The measured data for various proton energies is 
tabulated in  Table~\ref{tab:96g+mTc} together with  the
corresponding correlation coefficients.
Figure~\ref{fig:96g+mTc_CS} shows the excitation function of the
 $^{96\text{g+m}}$Tc production. The  present data shows that the cross section decreases from
about 12 to 16 MeV and is nearly constant thereafter, which is similar to earlier  measurements 
(\cite{tarkanyi2012investigation}$-$\cite{elbinawi2019study}, ~\cite{bonardi2002thin}$-$\cite{takacs2015reexamination}) .  
The improved precision and consequently a better quality of the data is evident from the figure.

\begin{figure*}[ht!]
\centering
\includegraphics[width=0.85\textwidth]{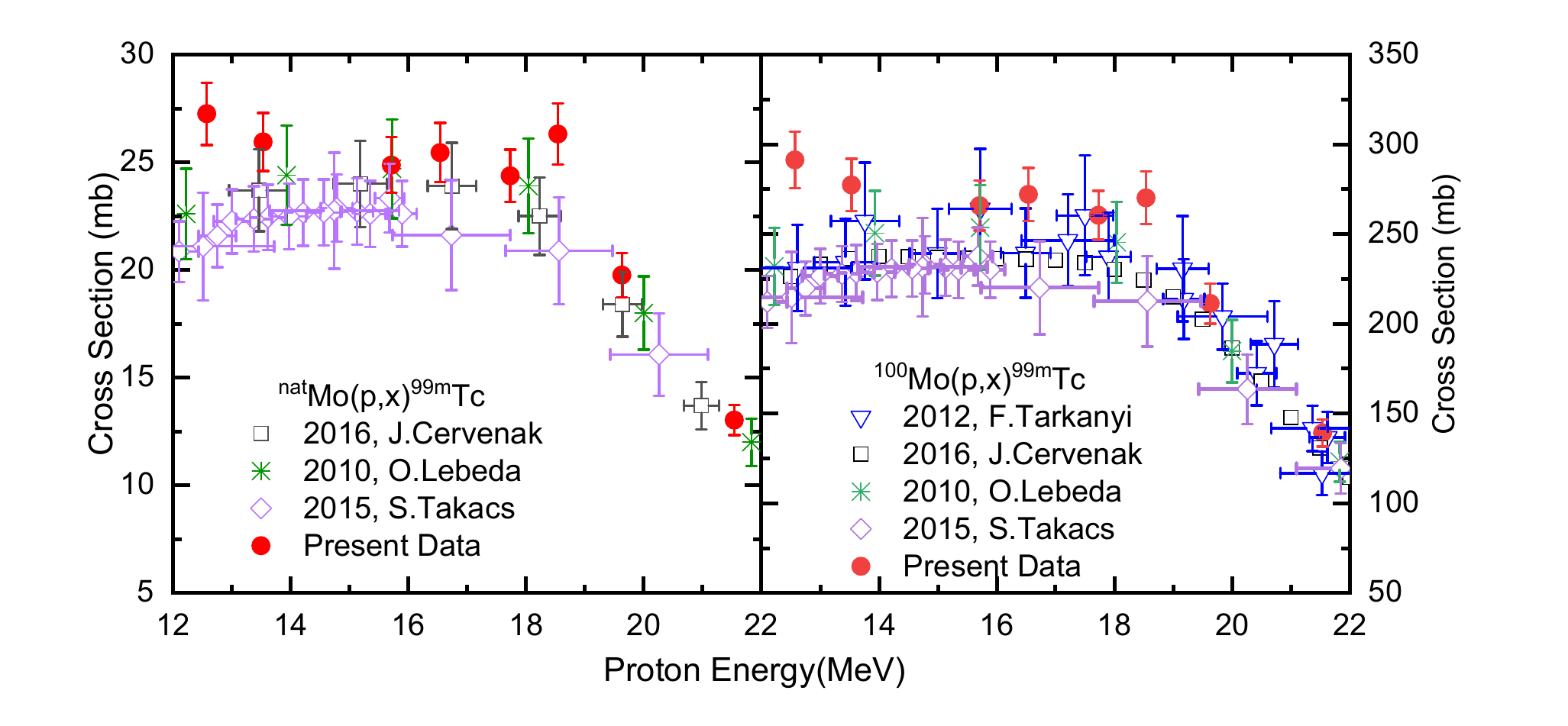}
\vspace{0.5cm} 
\caption{Excitation function for (a) $^{nat}$Mo(p,x)$^{99m}$Tc and (b) $^{100}$Mo(p,x)$^{99m}$Tc reactions.}
\label{fig:99mTc_CS}
\vspace{0.5cm} 
\end{figure*}

\begin{table*}[t]
\centering
 \caption{cross sections for $^{nat}$Mo(p,x)$^{99m}$Tc reaction together with fractional 
uncertainty in parameters and correlation coefficients between $E_p$ and $\sigma$.}
    \begin{tabular}{cccccccccccccccc}
    \hline
   E$_p$  & $\sigma$  & $\Delta \epsilon$ & $\Delta I_{\gamma}$  & $\Delta N_t $ & $\Delta N_{\gamma}$ & Total & \multicolumn{8}{c}{Correlation coefficients}\\
     (MeV) & (mb)&(\%) &(\%) &(\%) &(\%) & Uncertainty (\%) & \multicolumn{8}{c}{}\\
    \hline
21.54 & 13$\pm$1 & 2.16 & 4.49 & 0.52 & 1.98 & 5.39 & 1 &  &  &  &  &  &  &  \\
19.63 & 20$\pm$1 & 2.20 & 4.49 & 0.44 & 1.46 & 5.23 & 0.72 & 1 &  &  &  &  & &    \\
18.54 & 26$\pm$1 & 2.16 & 4.49 & 0.64 & 1.94 & 5.39 & 0.86 & 0.72 & 1 &  &  &  &  &   \\
17.73 & 24$\pm$1 & 1.57 & 4.49 & 0.44 & 1.30 & 4.95 & 0.76 & 0.91 & 0.76 & 1 &  &  &  &   \\
16.54 & 25$\pm$1 & 2.16 & 4.49 & 0.59 & 1.96 & 5.39 & 0.86 & 0.72 & 0.86 & 0.76 & 1 &  &  &   \\
15.71 & 25$\pm$1 & 2.20 & 4.49 & 0.46 & 1.30 & 5.19 & 0.72 & 0.92 & 0.72 & 0.92 & 0.72 & 1 &  & \\
13.53 & 26$\pm$1 & 2.20 & 4.49 & 0.53 & 1.38 & 5.22 & 0.72 & 0.92 & 0.72 & 0.91 & 0.72 & 0.92 & 1 &   \\
12.57 & 27$\pm$2& 2.16 & 4.49 & 0.55 & 1.79 & 5.33  & 0.87 & 0.72 & 0.87 & 0.77 & 0.87 & 0.73 & 0.73 & 1  \\
       \hline
    \end{tabular}
    \label{tab:natMo(p,x)99mTc}
\end{table*}

\begin{table*}[t]
\centering
\caption{cross sections for $^{100}$Mo(p,x)$^{99m}$Tc reaction together with fractional 
uncertainty in parameters and correlation coefficients between $E_p$ and $\sigma$.}
\begin{tabular}{cccccccccccccccc}
\hline
E$_p$  & $\sigma$  & $\Delta \epsilon$ & $\Delta I_{\gamma}$ & $\Delta a$ & $\Delta N_t $ & $\Delta N_{\gamma}$ & Total & \multicolumn{8}{c}{Correlation coefficients}\\
(MeV) & (mb)&(\%) &(\%) & (\%) &(\%) &(\%) & Uncertainty (\%) & \multicolumn{8}{c}{}\\
\hline
21.54 & 139$\pm$8 & 2.16 & 4.49 & 0.67 & 0.52 & 1.98 & 5.43 & 1 &  &  &  &  &  &  &  \\
19.63 & 211$\pm$11& 2.20 & 4.49 & 0.67 & 0.44 & 1.46 & 5.27 & 0.72 & 1 &  &  &  &  & &    \\
18.54 & 270$\pm$15& 2.16 & 4.49 & 0.67 & 0.64 & 1.94 & 5.43 & 0.86 & 0.72 & 1 &  &  &  &  &   \\
17.73 & 260$\pm$14& 2.20 & 4.49 & 0.67 & 0.44 & 1.30 & 5.23 & 0.73 & 0.92 & 0.73 & 1 &  &  &  &   \\
16.54 & 272$\pm$15& 2.16 & 4.49 & 0.67 & 0.59 & 1.96 & 5.43 & 0.86 & 0.72 & 0.86 & 0.73 & 1 &  &  &   \\
15.71 & 266$\pm$14 & 2.20 & 4.49 & 0.67 & 0.46 & 1.30 & 5.23& 0.73 & 0.92 & 0.73 & 0.93 & 0.73 & 1 &  & \\
13.53 & 277$\pm$15& 2.20 & 4.49 & 0.67 & 0.53 & 1.38 & 5.26 & 0.72 & 0.92 & 0.72 & 0.93 & 0.72 & 0.93 & 1 & \\
12.57 & 291$\pm$16& 2.16 & 4.49 & 0.67 & 0.55 & 1.79 & 5.37 & 0.87 & 0.73 & 0.87 & 0.73 & 0.87 & 0.73 & 0.73 & 1  \\
\hline
\end{tabular}
\label{tab:100Mo(p,x)99mTc}
\end{table*}

\subsection{Production of \(^{99}\)Mo:}
The main contributing channels to  $^{99}$Mo production are summarized in 
Table~\ref{tab:nndc-data}. Measured cross sections for the $^{nat}$Mo(p,x)$^{99}$Mo 
 are listed in 
Table~\ref{tab:natMo(p,x)99Mo} together with correlation coefficients and are plotted in 
Figures~\ref{fig:99Mo_CS}~(a). The derived cross section for and $^{100}$Mo(p,x)$^{99}$Mo reaction from the above data  are given in Table~\ref{tab:100Mo(p,x)99Mo} and are plotted in   in Figure~\ref{fig:99Mo_CS}(b). 
As can be seen from the Figure, the present data 
are in good agreement with prior studies
~\cite{khandaker2007measurement,lebeda2010new,chodash2011measurement,lamere2019proton,levkovski1991cross}.  
For $^{99}$Mo, the uncertainties in the present measurements  are  comparable to those reported in previous works. However, it should be pointed out that Levkovski et al.  ~\cite{levkovski1991cross} did not report the proton beam energy spread. 

\subsection{Production of \(^{99m}\)Tc:} 
The $^{99\text{m}}$Tc can be produced  via $^{100}$Mo(p,x) reactions as detailed in
Table~\ref{tab:nndc-data}. The contribution from $^{98}$Mo(p,$\gamma$) channel is also suggested by
~\cite{khandaker2007measurement}. The extraction of production cross section $^{99\text{m}}$Tc is
complicated due the contribution from  beta decay of $^{99}$Mo. The  $^{99}$Mo
predominantly decays to  the $^{99\text{m}}$Tc ($f\sim
87.9\%$)~\cite{matsuzaki202499mo}, a small fraction  (5.1\%)~\cite{takacs2015reexamination} decays to the
140.5 keV state in $^{99}$Tc and remaining fraction decays to the ground state of  $^{99}$Tc.
Hence, the observed photopeak yield of 140.5 keV line, $N_{140}$ needs to be corrected for the
beta decay contribution of $^{99}$Mo.
$$N_{140}^{m}=N_{140}^{obs}- N_{140}^{Mo}$$

Similar to Eq.~\ref{eq:N1362_counts}, $N_{140}^{m}$ can be written as

\begin{equation}
N_{140}^{m} = \varepsilon_{140}\, I_{\gamma ^1}
\int_{t_1}^{t_2}
\frac{d}{dt}
\left[
Y_{mTc}^{\mathrm{direct}}(t) + Y_{Mo \rightarrow m}(t)
\right] \, dt
\label{eq:N140_counts}
\end{equation}
where $\varepsilon_{140}$ and  $I_{\gamma 1}$ are photopeak efficiency and branching ratio of 140 keV gamma ray, respectively. 
The $^{99\mathrm{m}}$Tc production cross
section is obtained using the similar methodology described in Sections  4.2.3. 
The   yields of $^{99}$Mo and $^{99\mathrm m}$Tc from $^{99}$Mo decay at the end of the $i^{th}$ time interval, $Y_{\mathrm{Mo}}(i)$ and $Y_{\mathrm{Mo} \rightarrow m}$ (i), respectively,    can be written as
\begin{equation}
Y_{Mo}(i)
= Y_{Mo}(i-1)
+ P_{99Mo}(i)
- \lambda_1\,Y_{Mo}(i-1)\,dt(i)
\label{eq:99Mo_yield}
\end{equation}

\begin{equation}
\begin{split}
Y_{{Mo} \rightarrow m} (i) =
Y_{{Mo} \rightarrow m} (i-1) 
+ 0.879\lambda_1 Y_{{Mo}}(i-1) dt(i) \\
- \lambda_2 Y_{{Mo} \rightarrow m}(i-1) dt(i)
\end{split}
\label{eq:99mTc_Mo-yield}
\end{equation}
where  $\lambda_1$ and $\lambda_2$ are decay constants of 
$^{99}$Mo and  $^{99\mathrm m}$Tc, respectively and  $P_{99Mo}(i)$ is the production rate given by eq.~\ref{eq:P1}.
Although $\lambda_1 <<\lambda_2$, interval wise 
computation process has been adopted for better precision.  The contribution from $^{99}$Mo to $^{99m}$Tc decay in the observed photopeak yield of 140 keV, $N_{140}^{Mo\rightarrow m}$, is computed using the cross section $\sigma(^{99}Mo)$ obtained in Section 4.6.
Then the  contribution to the photopeak counts from the direct isomer production $^{99m}$Tc is  obtained as
\begin{equation}
    N_{140}^{m-direct} = N_{140}^m-  N_{140} ^{Mo\rightarrow m}
\end{equation}
The direct isomer production cross section $\sigma(^{99m}$Tc) is extracted from $N_{140}^{m-direct}$.



In Fig.~\ref{fig:99mTc_CS}~(a) and (b), the measured excitation functions of the $^{nat,100}$Mo(p,x)$^{99\mathrm m}$Tc reactions are compared
with previously reported data~\cite{tarkanyi2012investigation,vcervenak2016experimental,lebeda2010new,takacs2015reexamination}.
The observed trend in present cross sections is consistent 
with the reported values, namely, constant up to 18 MeV and rapid decrease thereafter. However, the absolute values of the measured cross section is higher at $E< 14$ MeV. 
It should be mentioned that the data for the $^{nat}$Mo(p,x)$^{99m}$Tc reaction reported  in 
~\cite{khandaker2007measurement} and ~\cite{lagunas1991cyclotron} 
correspond  only to the reaction 
$^{100}$Mo(p,2n)$^{99m}$Tc  and therefore are not directly comparable with our data. 
The measured cross sections  along with the corresponding correlation
coefficients are listed in Tables~\ref{tab:natMo(p,x)99mTc} 
and
\ref{tab:100Mo(p,x)99mTc}, respectively, for  different incident proton energies.

\section{Conclusion}
This study reports new measurements of $^{nat}$Mo(p,x) and $^{100}$Mo(p,x) reaction 
cross sections in the proton energy range of 12 to 22~MeV with substantially reduced uncertainties, as compared to previously reported data.
In the present work, the enhanced precision was achieved through usage of thin targets, precise
beam current measurements, and systematic corrections for 
isomeric contributions. 
For proton induced reactions on molybdenum, covariance information is reported for
the first time and provides correlation coefficients between the measured cross 
sections. Consequently, the present measurements are important not only for reduced uncertainties but also
for more precisely quantified error propagation, thereby providing greater
confidence in their application to nuclear data evaluation. Hence, these results 
contribute to strengthening international databases such as EXFOR and serve as 
reliable benchmarks for widely used nuclear reaction codes including TALYS, EMPIRE, 
and ALICE.
Equally important, the data also hold direct relevance for medical 
isotope production. The measured excitation functions for $^{99m}$Tc and $^{99}$Mo 
support their established use in SPECT imaging and generator systems, while 
$^{96g}$Tc and $^{95g}$Tc show promise as potential alternatives or tracers. 
Additional isotopes such as $^{94g}$Tc, $^{93g}$Tc and $^{89g}$Tc are of growing 
interest in PET imaging and targeted oncology. 
By combining reduced uncertainties with clinically relevant 
outcomes, this work strengthens the bridge between nuclear data research and 
healthcare applications, ensuring lasting impact on both nuclear science and 
medical diagnostics.

\section{Acknowledgement}
We thank the PLF  staff for  the smooth accelerator operation, Mr. R.D. Turbhekar and Mr.
N.C. Kamble for the target preparations,  Mr. Nishant Jangid and Mr. 
Kiran Divekar for assistance during the experiment and N.Otsuka for his guidance during analysis. The  financial support 
from the  Science and Engineering Research Board, Government of India (GoI)(Sanction 
No. EEQ/2022/000545, dated 17.02.2023), from the Institution of Eminence (IoE), Banaras
Hindu University (Grant No. 6031-B) and from the Department of Atomic Energy (GoI), 
under Project No. RTI4002 is gratefully acknowledged. One of the authors, SB, acknowledges the 
financial support provided by the University Grants Commission (UGC), New Delhi, India, under the
Junior Research Fellowship (JRF) scheme [fellowship award reference number:231610089402].

\end{document}